\documentclass[fleqn,usenatbib,useAMS]{mnras}

\usepackage{graphicx}	
\usepackage{amsmath}	
\usepackage{amssymb}	
\usepackage{multicol}        
\usepackage{bm}		
\usepackage{pdflscape}	

\usepackage[T1]{fontenc}
\usepackage{ae,aecompl}

\usepackage{newtxtext,newtxmath}

\title[Photometric and Orbital Period Investigation of TYC 4002-2628-1]{The First Photometric and Orbital Period Investigation of an Extremely Low Mass Ratio Contact Binary with a Sudden Period Change, TYC 4002-2628-1}

\author[D. F. Guo et al.]{
Di-Fu. Guo,$^{1}$\thanks{E-mail: difu@sdu.edu.cn}
Kai Li,$^{1}$\thanks{E-mail: kaili@sdu.edu.cn}
Fen Liu,$^{1}$ 
Huai-Zhen Li$^{2}$
Qi-Qi Xia,$^{1}$
Xing Gao,$^{3}$
Xiang Gao,$^{1}$
Xu Chen,$^{1}$ 
\newauthor
Dong-Yang Gao,$^{1}$
and Guo-You Sun$^{4}$ \\
$^{1}$Shandong Provincial Key Laboratory of Optical Astronomy and Solar-Terrestrial Environment, Institute of Space Sciences, Shandong University, Weihai \emph{}264209, China\\
$^{2}$Department of Physics, Yuxi Normal University, Yuxi, Yunnan, 653100, China\\
$^{3}$Xinjiang Astronomical Observatory, 150 Science 1-Street, Urumqi 830011, China\\
$^{4}$Wenzhou Astronomical Association
}

\date{Accepted XXX. Received YYY; in original form ZZZ}

\pubyear{2022}

\begin{document}
\label{firstpage}
\pagerange{\pageref{firstpage}--\pageref{lastpage}}
\maketitle

\begin{abstract}
Photometric observations for the totally eclipsing binary system TYC 4002-2628-1, were obtained between November 2020 and November 2021. To determine the stellar atmospheric parameters, a spectral image was taken with the 2.16 m telescope at National Astronomical Observatory of China (NAOC).  TYC 4002-2628-1 is a low-amplitude (about 0.15 mag for $V$ band), short-period (0.3670495 d), contact eclipsing binary with  a total secondary eclipse. Intrinsic light curve variations and the reversal of the O'Connell effect are detected in the light curves, which  may be due to spot activity. 
Based on the photometric solutions derived from the multi-band time series light curves, TYC 4002-2628-1 is an extremely low mass ratio contact binary with a mass ratio of  $q\sim$ 0.0482 and a fill-out factor of $f\sim5\%$. By analyzing the $O-C$ variations, we find that its orbital period remains unchanged when  BJD < 2458321 . Then the orbital period changed suddenly around  BJD 2458743 and  has an increasing rate of
$dP/dt=1.62\times{10^{-5}}day\cdot yr^{-1}=140$ $second\cdot century^{-1}$ .
 If confirmed, TYC 4002-2628-1 would be the contact binary with the highest orbital period increasing rate so far.  By investigating the ratio of orbital angular momentum to the spin angular momentum ( $J_{orb}$/$J_{spin}$ $<3$) , the instability mass ratio ($q_{inst}/q=1.84$) and the instability separation ($A_{inst}/A=1.35$),  TYC 4002-2628-1 can be regarded as a merger candidate.
\end{abstract}




\begingroup
\let\clearpage\relax
\tableofcontents
\endgroup
\newpage

\section{Introduction}

W UMa binaries are short-period and low-temperature binary systems, with the two components over-filling their respective critical Roche lobe and sharing a common envelope. At the same time, the light curves are generally asymmetric and the maxima are usually unequal. The difference of the maximum values of the light curve, often caused by the strong magnetic activity (star spot) due to the deep convective envelopes, is usually referred to O'Connell effect \citep{OConnell1951}.  Among W UMa binaries, deep (degree of overcontact $f > 50.0 \%$), low mass ratio (mass ratio $q < 0.25$) overcontact binary systems play quite a critical role for studying and understanding the dynamical evolution of binaries since they are at late evolutionary stage, and may merge into rapidly rotating stars \citep{qian2005a}. Such stellar mergers are very rare, occurring about once every decade in our galaxy \citep{Kochanek2014}. So far, only one binary merging event, V1309 Sco \citep{tylenda2011, zhu2016}, has been found. Searching for such merger candidates of contact binary is quite important for studying stellar astrophysics.

Theoretical study indicates that a tidal instability would happen (Darwin's instability) in a contact binary system when the mass ratio is less than the theoretical limit,  which ultimately drives the system to evolve into a single, rapidly rotating object \citep{Hut1980}. If this is the case, it may imply that such systems  would not be observed.  Based on different theoretical assumptions, the minimum mass ratio of contact binaries have been investigated by several authors.  \cite{Rasio1995a} suggested that the lowest mass ratio for contact binaries possessing two unevolved main sequence stars is $q_{min} \simeq 0.09$. \cite{Li2006} derived a smaller value of $q_{min}\simeq 0.076$ by supposing W UMa systems rigorously comply with the Roche geometry. \cite{Arbutina2007} found that the theoretical minimum mass ratio is ranged from 0.094 to 0.109. Later, \cite{Arbutina2009} reanalyzed the minimum mass ratio by considering rotating
polytropes and concluded $q_{min}$ can achieve 0.070-0.074.  On the basis of the primary structure and mass, \cite{Jiang2010} argued that the $q_{min}$ can be as lower as 0.05.  Up to now, only a few binaries have been found that the mass ratio is below 0.090, e.g., V1187 Herculis ($q \sim 0.044 $, \citealt{caton2019}),  VSX J082700.8$+$462850 ($q \sim 0.055 $, \citealt{Li2021a}), V857 Her ($q \sim 0.065 $, \citealt{qian2005}), SX Crv ( $q \sim 0.072 $, \citealt{zola2004}), ASAS J083241+2332.4 ($q \sim 0.068 $, \citealt{Sriram2016}), ZZ PsA ($q \sim 0.078 $, \citealt{Wadhwa2021}), M4 V53 ($q \sim 0.078 $, \citealt{Li2017}),  NSV 13890 ($q \sim 0.080 $, \citealt{Wadhwa2006}),  V870 Ara ($q \sim 0.082 $, \citealt{szalai2007}). Owing to their extremely low mass ratios around the theoretical limit, the discovery of such systems plays an important role in testing and constraining the theoretical evolutionary mode concerning binary mergers. However, due to the faintness of the less massive star, radial velocity (RV) determination for the extremely low mass ratio binaries is quite difficult.  Total eclipsing binaries provide another channel to obtain reliable photometric solutions for the extremely low mass ratio systems without the need of the RV observations \citep{Terrell05,Li2021b}.

W UMa stars are customarily classified into two subclasses: A-type and W-type \citep{Binnendijk1970}, according to the shape of the light curve. For A-type binaries, the primary minimum is caused by the transit of the less massive component. On the contrary, the primary minimum of W-btype binaries results from the occultation of the more massive component. However, some binaries, e.g., V857 Her \citep{qian2005a}, V802 Aql \citep{samec2004}, and RT LMi \citep{Qian2008}  reveal A-type light curves, but the photometric solutions show that the less massive components are hotter than the more massive ones, which imply W-type systems. In addition, a few contact binaries have changed their types among A and W at an interval of several years, like AM Leo (\citealt{Binnendijk1969,Hoffmann1982,Derman1991}), AH Cnc (\citealt{Maceroni1984,Zhang2005,Qian2006}), SS Ari ( \citealt{Kim2003, Kurochkin1960}) and FG Hya \citep{qian2005}. These indicate that the type of a binary cannot be distinguished based on the shape of the light curve alone.

 According to the International Variable Star Index \footnote{https://www.aavso.org/vsx/index.php}
 (VSX) and the SIMBAD website \footnote{http://simbad.cds.unistra.fr/simbad/}, the object  TYC 4002-2628-1 (other names: CzeV710, UCAC4 725-101725, WISE J230927.8+545123) was first discovered as a binary by \cite{Heinze2018} with an extremely small amplitude of 0.138 mag \citep{chen2020}. The orbital period of the target is 0.3670495 days. In this paper, the first complete $BVR_cI_c$ band observations of TYC 4002-2628-1, exhibiting a total eclipse, are presented in Section 2. In Section 3, the orbital period variations were investigated based on the light minimum times. In Section 4, the photometric solutions were provided. Conclusion and discussion are summarized in Section 5.

\section{Observations and data reduction}

\subsection{Photometric observations}

\begin{figure}
\begin{center}
\begin{tabular}{c@{\hspace{0.3pc}}c}
\includegraphics[width=8cm]{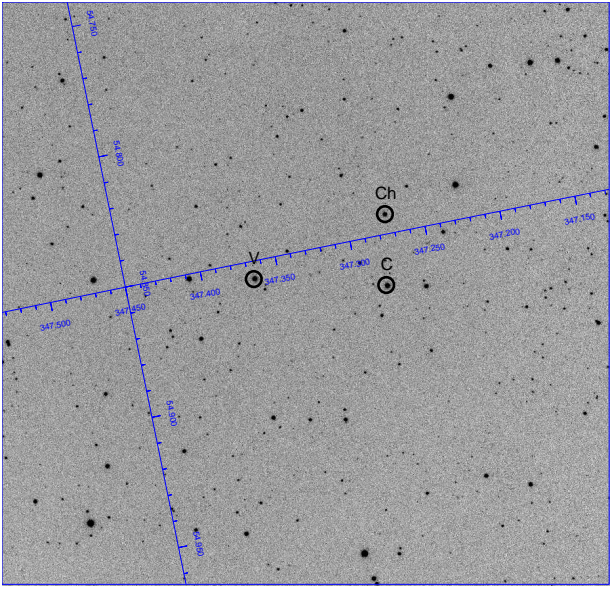}
\end{tabular}
\end{center}
\caption{ The middle area of the CCD image obtained by the Weihai 60cm telescope is used as the finding chart, TYC 4002-2628-1 (V), comparison (C) and check (Ch).  The field of view is about $12'$ $\times$ $12'$ }
\label{fig_chart}
\end{figure}

Photometric observations of the total eclipse binary system TYC 4002-2628-1 were carried out from November 2020 to November 2021 with the 0.6m Ningbo Bureau of Education and Xinjiang Observatory
Telescope (NEXT)  and the 0.6m reflecting telescope at Weihai Observatory (WH60). A back illuminated FLI 230-42 CCD camera was mounted to the NEXT telescope. The camera has $2048\times2048$  square pixels with a field of view of $22'$ $\times$ $22'$ . As for WH60, SBIG STX-16803 was used to obtain the CCD images. The CCD  possesses $4096\times4096$  square pixels with a field of view of  $30'$ $\times$ $30'$ .  The standard Johnson-Cousins filters ($B$, $V$, $R_c$, and $I_c$) were used during the observations.
During each observing night, bias frames, dark frames and twilight sky flats were observed at the beginning or at the end of the observation. Otherwise, adjacent flat frames were used in the subsequent correction. Based on the IRAF software, the differential light curves were derived from aperture photometry.  The finding chart and the basic information about the comparison star and the check star employed by us are exhibited in Figure \ref{fig_chart} and Table \ref{coor}, respectively.  The exposure time for the observation obtained by NEXT is 30 s for $V$, 8 s for $R_{c}$ and 12 s for $I_{c}$, respectively. While the typical exposure time for WH60 is 80 s for B, 25 s for $R_{c}$ and 20 s for $I_{c}$, respectively.

\begin{figure*}
\begin{center}
\begin{tabular}{c@{\hspace{0.3pc}}c}
\includegraphics[angle=0,scale=0.28]{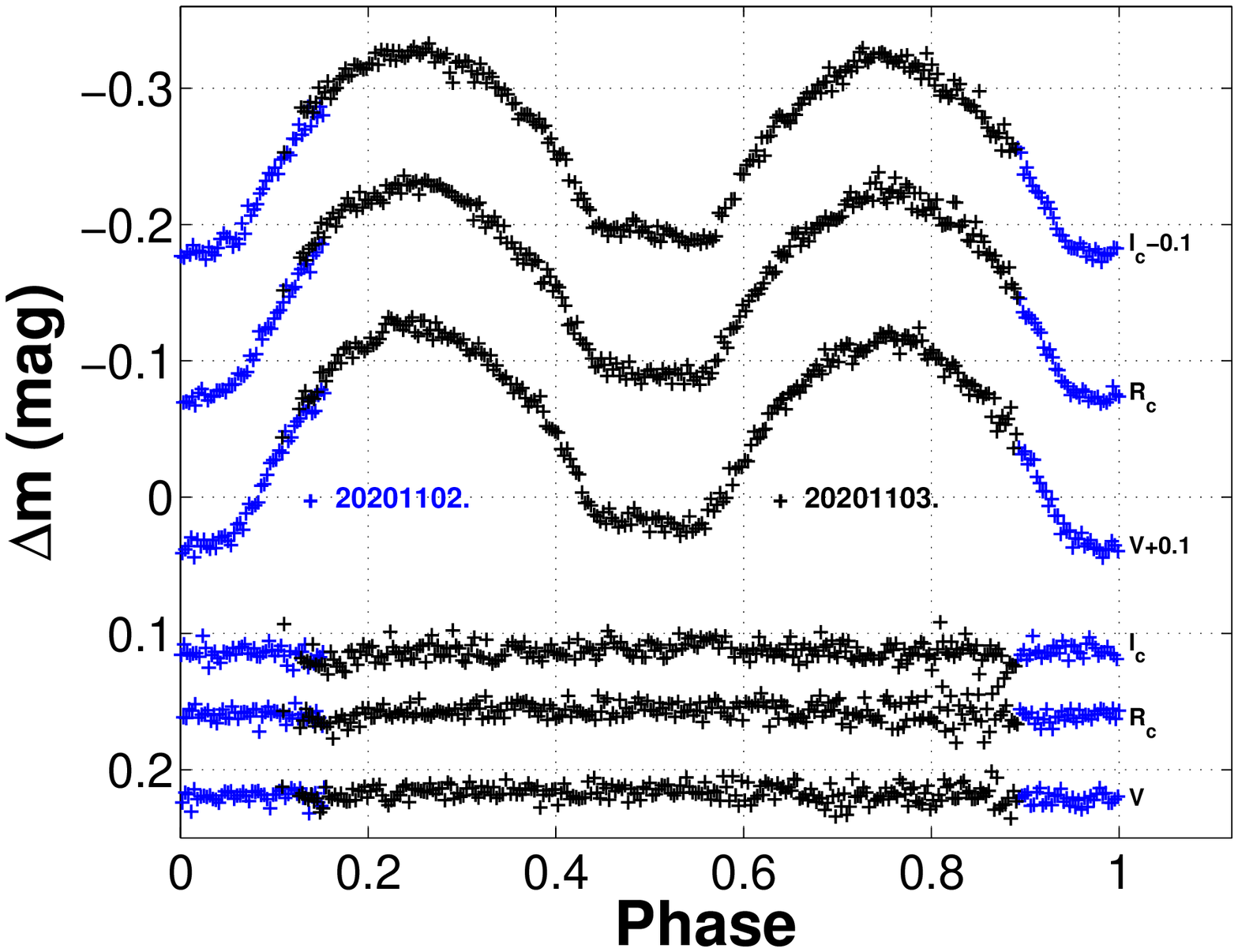}
\includegraphics[angle=0,scale=0.28]{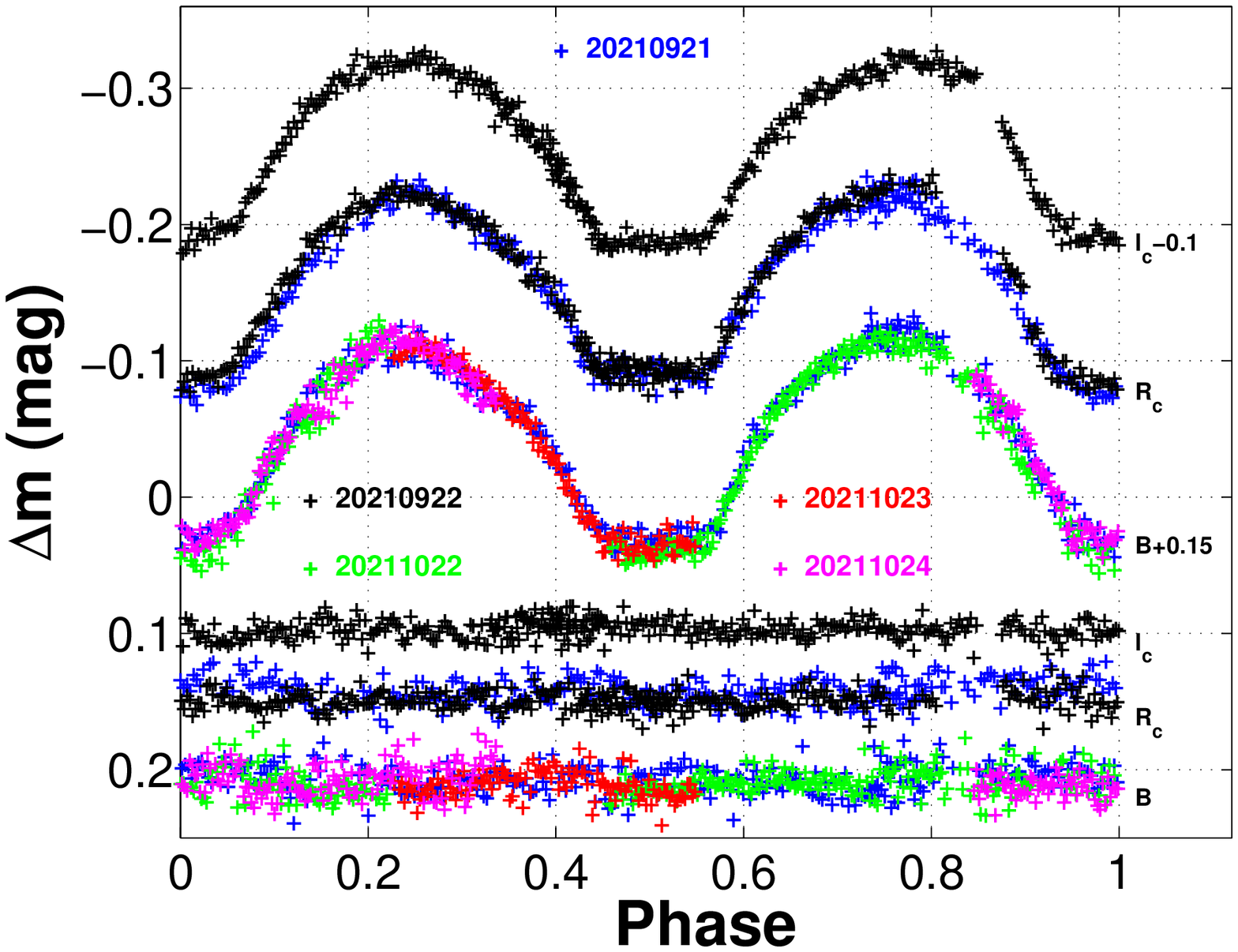}
\includegraphics[angle=0,scale=0.28]{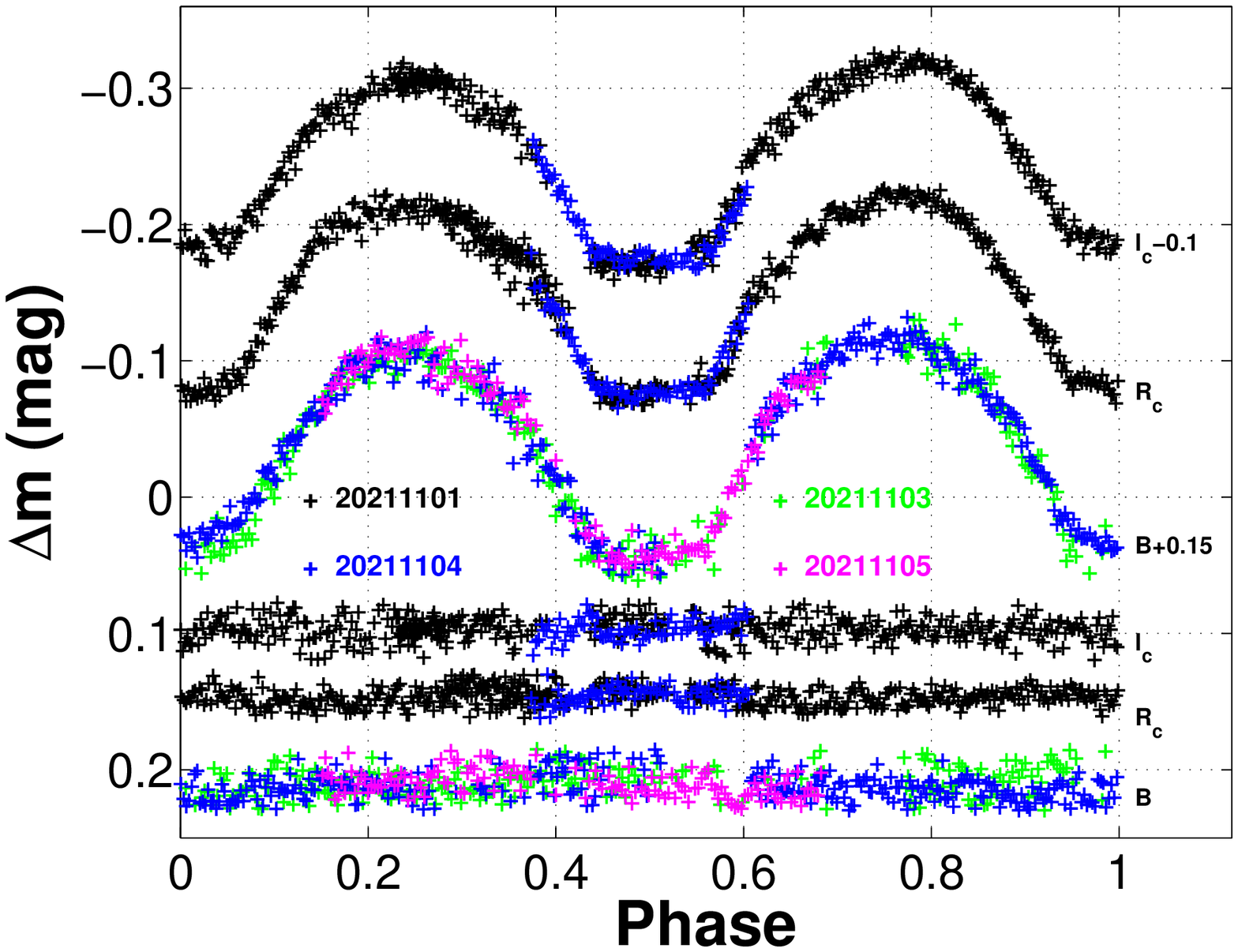}
\end{tabular}
\end{center}
\caption{Muti-band light curves of  TYC 4002-2628-1. The left panel is $LC_{1}$ observed in 2020 November, the middle panel is $LC_{2}$ observed in 2020 September and October, while the right panel is $LC_{3}$ obtained in 2021 November. The difference magnitudes of the comparison star and the check star are displayed at the bottom of the corresponding panel (constants added). }
\label{fig-all}
\end{figure*}

\begin{figure*}
\begin{center}
\begin{tabular}{c@{\hspace{0.3pc}}c}
\includegraphics[angle=0,scale=0.45]{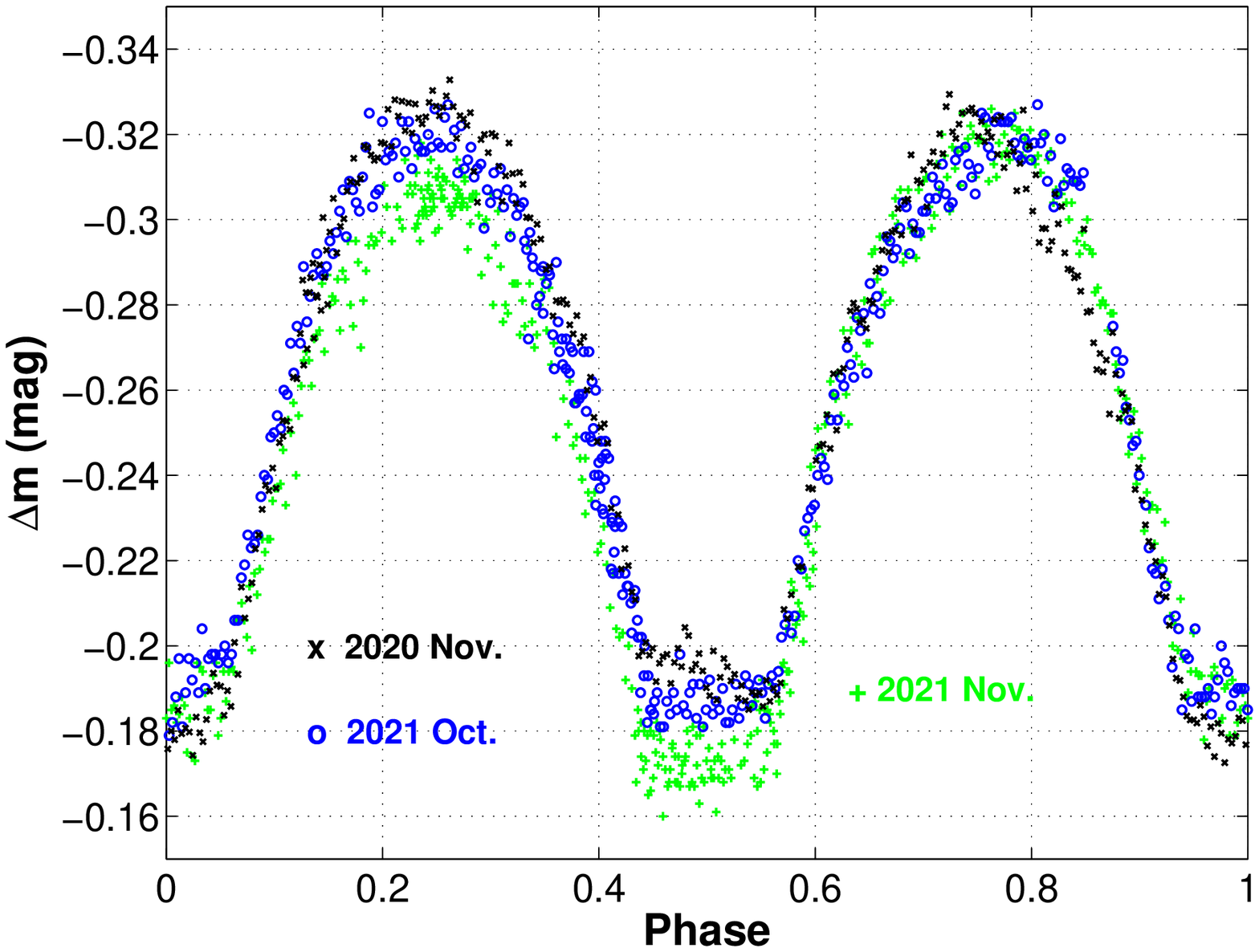}
\includegraphics[angle=0,scale=0.45]{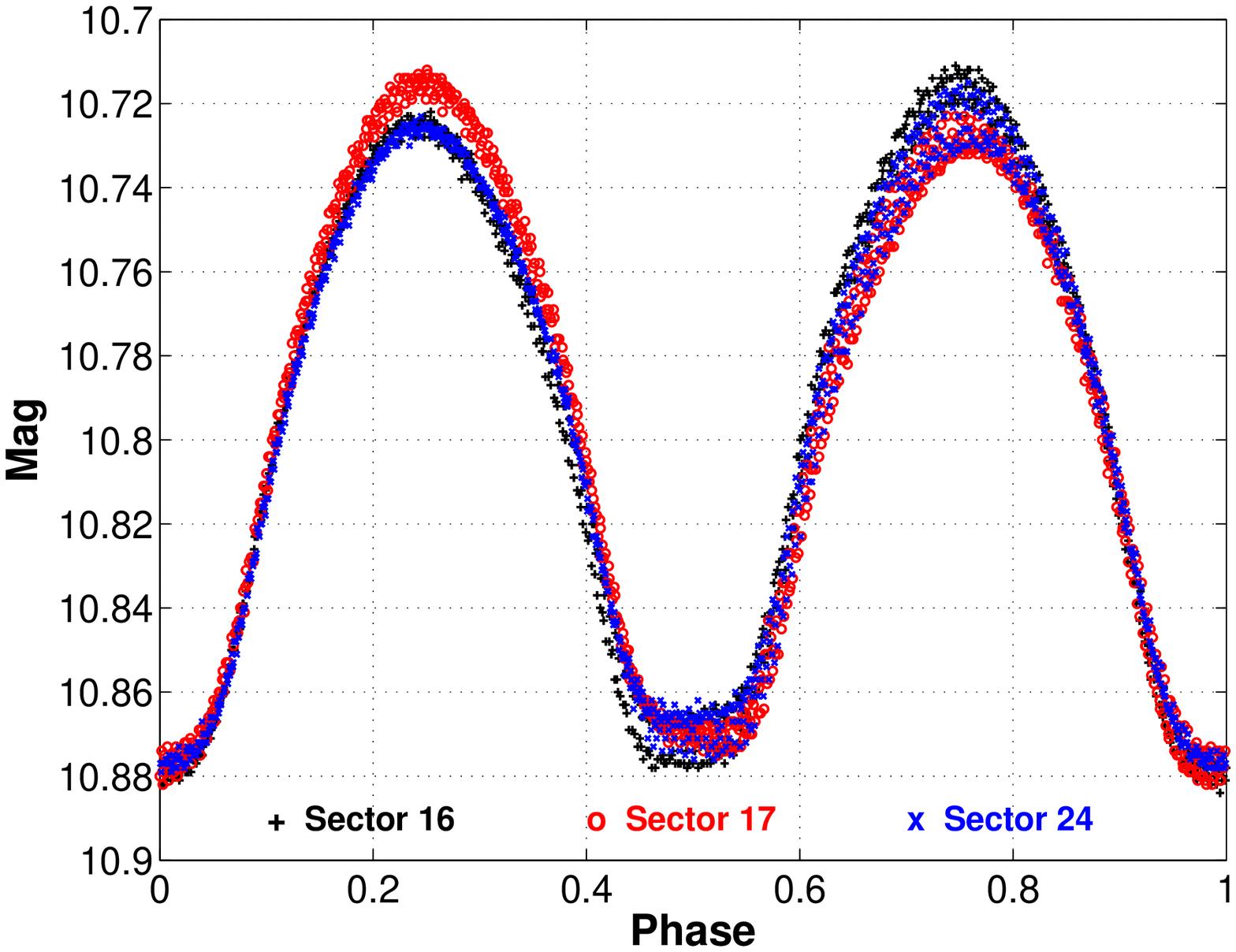}
\end{tabular}
\end{center}
\caption{Comparison of the light curves observed at different times. The left panel shows the light curves of $I_{c}$ band, while the right panel shows the light curves of TESS.}
\label{fig-compare}
\end{figure*}

To phase the light curves, the following ephemeris equation was used:
 \begin{eqnarray}
Min.I = HJD2459157.16588+ 0.3670495E.
\end{eqnarray}
According to the observation time and the shape of the light curves, three complected light curves ( e.g., light curve one ($LC_{1}$) in 2020 November, light curve two ($LC_{2}$) in 2021 September and October, light curve three ($LC_{3}$) in 2021 November) were obtained, which are shown in Figure \ref{fig-all}.  From Figure \ref{fig-all}, we can see that the light curves are varying with time, more or less, which may be caused by spot activities. To better compare the difference of light curve variations, the three $I$-band Light curves of TYC 4002-2628-1 are shown in the left panel of Figure \ref{fig-compare} as an illustration.  Based on Figure \ref{fig-compare}, the variations of the three sets of light curves are obvious, especially at primary maximum and secondary minimum. A positive O'Connell effect can be found in LC$_{1}$, while LC$_{3}$ shows an obvious negative O'Connell effect.  These indicate that the magnetic activity of TYC 4002-2628-1 is relatively active.

\begin{table*}
\begin{center}
\caption{Main parameters of TYC 4002-2628-1 and the reference stars.}\label{coor}
\begin{small}
\begin{tabular}{cccccc}\hline\hline
Targets          &   name               & $\alpha_{2000}$           &  $\delta_{2000}$   &  $V_{mag}^*$ &  $B-V$           \\ \hline
Variable(V)         &   TYC 4002-2628-1         & $23^{h}09^{m}27^{s}.87$    & $+54^\circ51'23''.5$   & $11.52$ & $0.72$             \\
The comparison (C)   &   GSC 0400202568  & $23^{h}09^{m}07^{s}.66$     & $+54^\circ52'08''.0$   & $11.70$ & $0.76$      \\
The check (Ch)        &   GSC 0400202542  & $23^{h}09^{m}05^{s}.87$    & $+54^\circ50'33''.0$   & $12.37$ & $0.56$     \\
\hline\hline
\end{tabular}
\end{small}
\end{center}
\textbf
{\footnotesize Note }
*  Magnitudes were taken from the catalog of APASS DR9  \citep{Henden2016}.
\end{table*}

We searched for the Transiting Exoplanet Survey Satellite (TESS; \citealt{ricker2015}) observation data of the target, and found it was observed during Sectors 16, 17, and 24 at 30-minute cadence. The Pre-search Data Conditioning Simple Aperture Photometry (PDCSAP) flux values ($F_{i}$) of the target can be derived from the Mikulski Archive for Space Telescopes
(MAST)\footnote{https://archive.stsci.edu/}. Then we converted the flux to magnitudes ($m_{i}$), using the following equation: $m_{i}= -2.5log(F_{i})$. The TESS magnitude of 10.78 derived from the fits head was used during the derivation of the magnitudes. Local eclipsing minima were used to fold the light curves, which are exhibited in the right panel of Figure \ref{fig-compare}. The intrinsic variations of the of light curves among different sectors are obvious, especially at the maxima and secondary minimum. A negative O'Connell effect can be found during Sectors 16 and then changes to  positive in Sector 17. While a reversal of the O'Connell effect was found in Sector 24.

\subsection{Spectral observation}

\begin{figure}
\begin{center}
\begin{tabular}{c@{\hspace{0.3pc}}c}
\includegraphics[width=8cm]{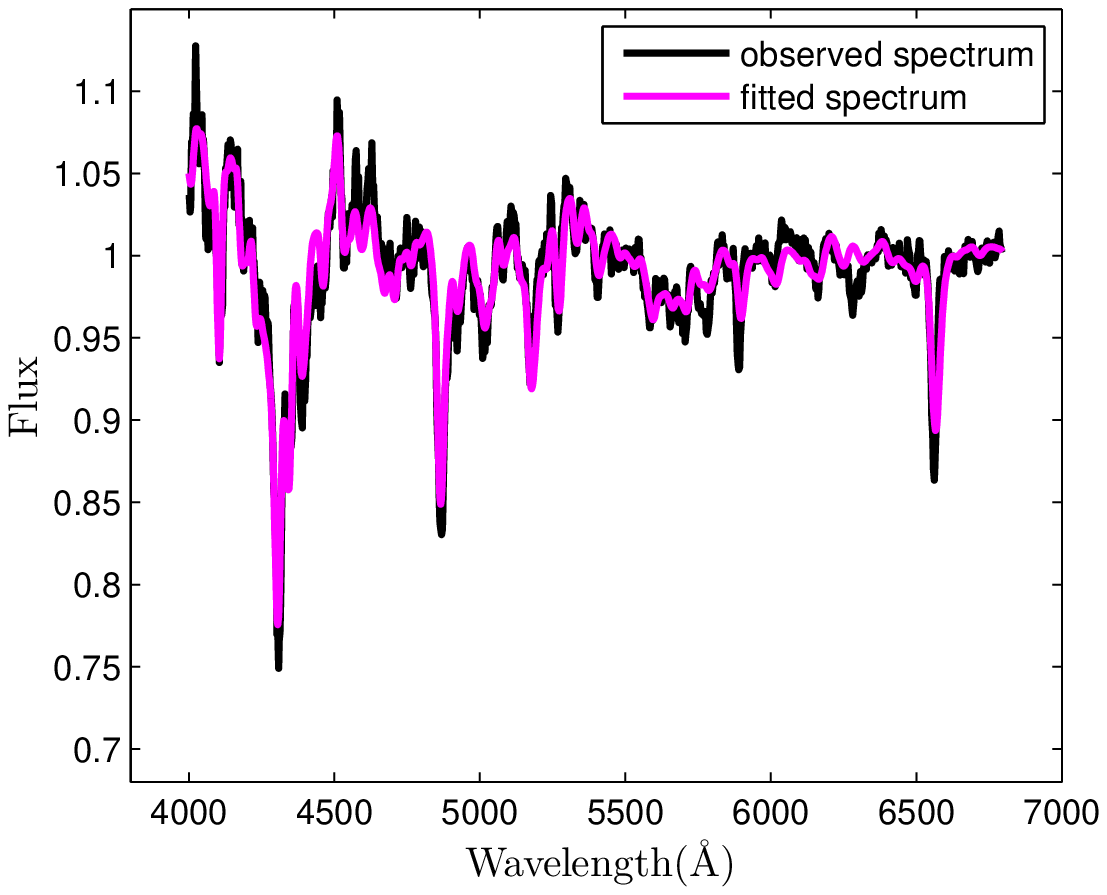}
\end{tabular}
\end{center}
\caption{ The observed and fitted spectrum of TYC 4002-2628-1. }
\label{fig_spectrum}
\end{figure}

To estimated the temperature of the primary star, on January 5, 2022, a spectral image of the target at phase 0.55 was taken by the Beijing Faint Object Spectrograph and Camera (BFOSC) mounted on the 2.16-m telescope at NAOC.
During the observation, low-dispersion spectrometer BFOSC and grism G4 were used, and the spectral resolution of a single pixel was 4.45 {\AA} \citep{Fan2016}.
The IRAF software was applied to process the
spectral data and extract the spectrum.
After that the University of Lyon Spectroscopic analysis Software (ULySS) \citep{Koleva2009} was applied to derive the atmospheric parameters of the system. The template spectra used for fitting are derived from an interpolator with the ELODIE library \citep{Prugniel2001}.
The normalized spectrum and the fitting spectrum are illustrated in Figure \ref{fig_spectrum}. The determined atmospheric parameters are as follows: $T_{eff} = 6032$ $\pm 26$ K, log$ g$= $4.3$ $\pm 0.06$ $cm s^{-2}$, [Fe/H] = 0.019$\pm0.04$ dex.

\section{Orbital Period Investigation}
\begin{figure*}
\begin{center}
\begin{tabular}{c@{\hspace{0.3pc}}c}
\includegraphics[angle=0,scale=0.45]{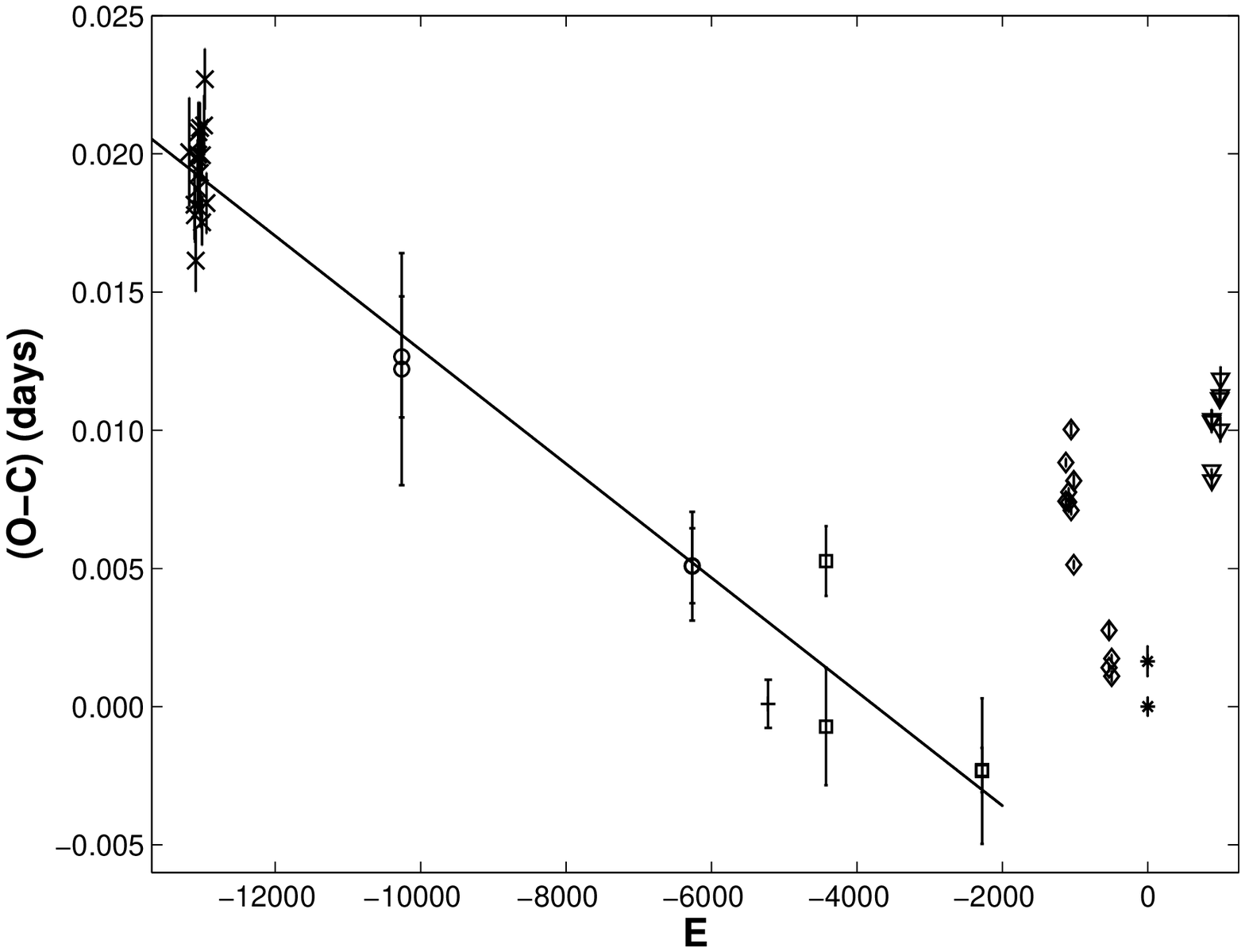}
\includegraphics[angle=0,scale=0.435]{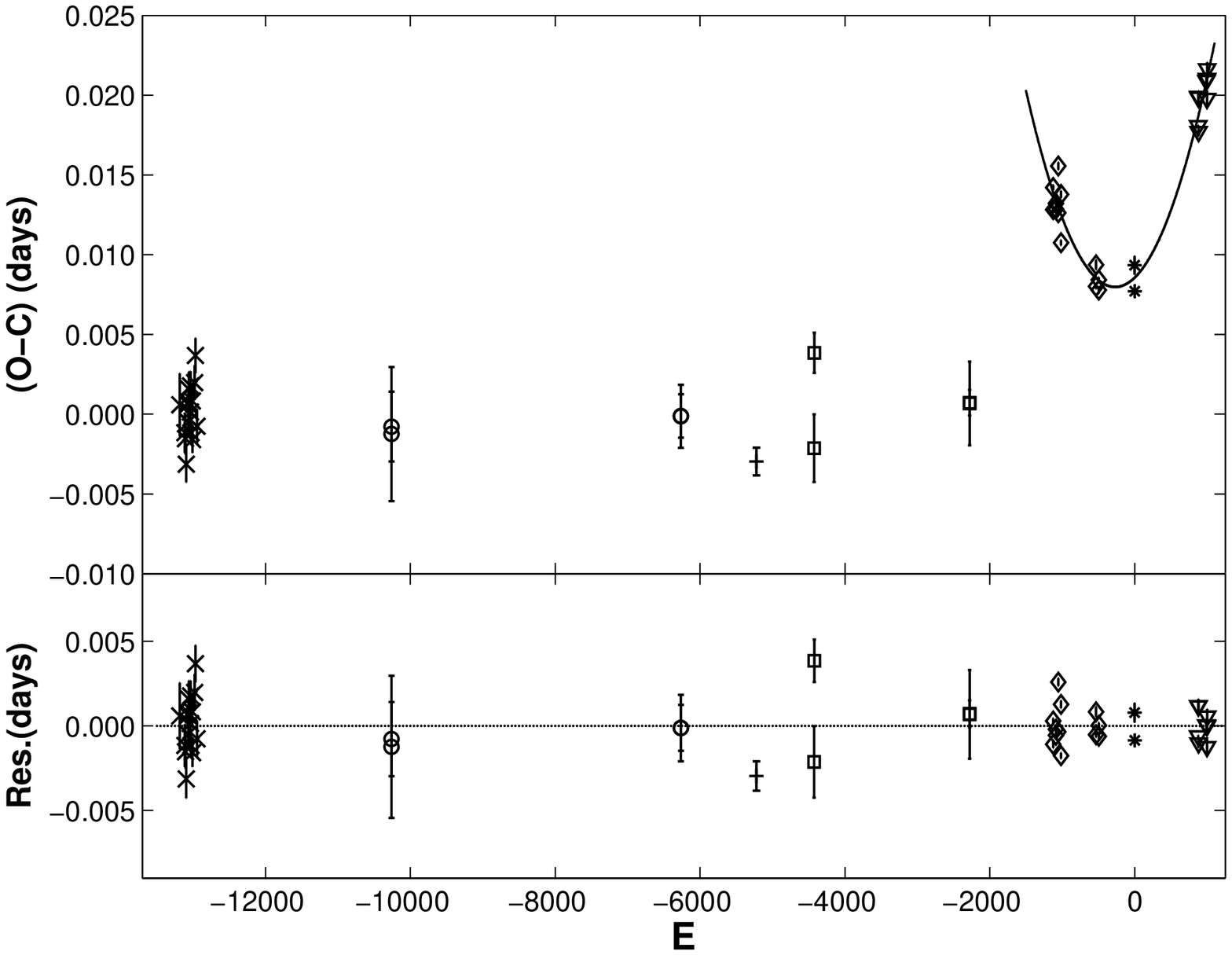}
\end{tabular}
\end{center}
\caption{ O-C diagram of TYC 4002-2628-1. x-mark, Superwasp; open circle, WISE;  plus, CzeV; square, ASAS-SN; diamond, TESS; star, NEXT; triangle, WH60.}
\label{oc}
\end{figure*}

 In order to study the orbital period variation of TYC 4002-2628-1, we use the optical sky survey data, i.e., Wide Angle Search for Planets (SuperWASP; \citealt{Butters2010}), 
 All-Sky Automated Survey for SuperNovae (ASAS-SN; \citealt{Shappee2014,Jayasinghe2018}), Wide-field Infrared Survey Explorer(WISE, \citealt{chen2018}) and the Transiting Exoplanet Survey Satellite (\citealt{ricker2015}),  to calculate its minimum times with the K-W method \citep{Kwee1956}. Due to the low time resolution of
WISE, ASAS-SN and TESS (30 min for TYC 4002-2628-1), we have used the method of \cite{Li2020,Li2021a} to computer the eclipsing minima. 46 new eclipsing minima were determined, which are illustrated in Table \ref{Newminimum}. In order to coincide with TESS time, HJD were transformed to BJD by using online tables\footnote{https://astroutils.astronomy.osu.edu/time/hjd2bjd.html} \citep{Eastman2010}. Then the following equation:
\begin{equation}
\begin{array}{lll}
Min.I(BJD) = 2459157.166669 + 0.3670495E
\end{array}
\label{oc1}
\end{equation}
was applied to calculate the $O - C$ values.
Based on Equation \ref{oc1}, the $O - C$ values are derived and illustrated in the left panel of Figure \ref{oc}. As shown from the left panel of Figure \ref{oc}, the $O - C$ values can be split into two parts. The first part of $O - C$ values (ranging from BJD 2454318 to 2458321) show an obvious linear variation, which may indicate the inaccuracy of the period. In order to correct this inaccuracy, the equation after the linear correction is as follows:
\begin{equation}
\begin{array}{lll}
\boldmath{ Min.I(BJD) = 2459157.158969(\pm0.002398) +}\\ \boldmath{0.3670474(\pm0.000001)E}
\end{array}
\label{oc2}
\end{equation}

The recalculated values of $ O - C $ derived from Equation \ref{oc2} are shown in the right panel of Figure \ref{oc}. From the right panel of Figure \ref{oc}, the second part of $O - C$ values (ranging from BJD 2458743 to 2459525) show a parabolic variation, and the following equation was obtained:
\begin{equation}
\begin{array}{lll}
 \boldmath{
 Min.I(BJD) = 2459157.167537(\pm0.001038) + }\\ \boldmath{0.3670518(\pm0.0000001)E +  8.124(\pm1.137)\times{10^{-9}}\times{E^{2}}}.
\end{array}
\label{oc3}
\end{equation}
The residuals are illustrated in the bottom of the right panel of Figure \ref{oc}, where no obvious variations were found.
The ephemeris shows a secular period increase which is determined to be  $dP/dt=1.62\times{10^{-5}}day\cdot yr^{-1}$. The black line in Figure \ref{oc} refers to the fitting curve derived from the quadratic fit.

\begin{table*}
\begin{center}
\caption{CCD times of light minimum for TYC 4002-2628-1.}
\label{Newminimum}
\scriptsize
\begin{tabular}{cccccccccccccc}\hline

HJD +2450000      & BJD +2450000   &  Error (days)     & Min.    & E      & O-C(day)          &  HJD+2450000      & BJD +2450000     &      Error (days)   & Min. & E      & O-C(day)                \\\hline
4318.55593$^a$    & 4318.556693    & $\pm0.00197 $    &   II    & -13182.5     &   0.00059          & 8320.84124$^d$    & 8320.842050      & $\pm0.00263 $        &  II     & -2278.5   &   0.00068         \\  
4344.61455$^a$    & 4344.615313    & $\pm0.00126 $    &   II    & -13111.5     &  -0.00116          &                   & 8743.87776$^e$   & $\pm0.00015 $        &  I      & -1126     &   0.01421     \\       
4347.55056$^a$    & 4347.551323    & $\pm0.00100 $    &   II    & -13103.5     &  -0.00153          &                   & 8744.05989$^e$   & $\pm0.00022 $        &  II     & -1125.5   &   0.01282         \\   
4351.58647$^a$    & 4351.587233    & $\pm0.00113 $    &   II    & -13092.5     &  -0.00314          &                   & 8757.45752$^e$   & $\pm0.00015 $        &  I      & -1089     &   0.01322     \\       
4360.58289$^a$    & 4360.583653    & $\pm0.00067 $    &   I     & -13068       &   0.00062          &                   & 8757.64069$^e$   & $\pm0.00023 $        &  II     & -1088.5   &   0.01286         \\   
4361.49939$^a$    & 4361.500153    & $\pm0.00081 $    &   II    & -13065.5     &  -0.00050          &                   & 8769.93655$^e$   & $\pm0.00017 $        &  I      & -1055     &   0.01263     \\       
4363.51941$^a$    & 4363.520172    & $\pm0.00091 $    &   I     & -13060       &   0.00076          &                   & 8770.12300$^e$   & $\pm0.00022 $        &  II     & -1054.5   &   0.01556         \\   
4364.43785$^a$    & 4364.438612    & $\pm0.00109 $    &   II    & -13057.5     &   0.00158          &                   & 8783.51542$^e$   & $\pm0.00016 $        &  I      & -1018     &   0.01075     \\       
4371.59384$^a$    & 4371.594602    & $\pm0.00151 $    &   I     & -13038       &   0.00014          &                   & 8783.70197$^e$   & $\pm0.00022 $        &  II     & -1017.5   &   0.01378         \\   
4372.51309$^a$    & 4372.513852    & $\pm0.00092 $    &   II    & -13035.5     &   0.00177          &                   & 8961.16365$^e$   & $\pm0.00014 $        & I       & -534      &  0.00802      \\       
4374.52892$^a$    & 4374.529682    & $\pm0.00067 $    &   I     & -13030       &  -0.00116          &                   & 8961.34852$^e$   & $\pm0.00029 $        & II      & -533.5    &  0.00937          \\   
4381.50481$^a$    & 4381.505572    & $\pm0.00095 $    &   I     & -13011       &   0.00083          &                   & 8974.74480$^e$   & $\pm0.00015 $        & I       & -497      &  0.00842      \\       
4382.42001$^a$    & 4382.420772    & $\pm0.00083 $    &   II    & -13008.5     &  -0.00159          &                   & 8974.92769$^e$   & $\pm0.00020 $        & II      & -496.5    &  0.00778          \\   
4393.43500$^a$    & 4393.435762    & $\pm0.00109 $    &   II    & -12978.5     &   0.00198          & 9156.24990$^f$    & 9156.250689      & $\pm0.00054 $        & II      &   -2.5    &  0.00934          \\  
4397.47422$^a$    & 4397.474982    & $\pm0.00110 $    &   II    & -12967.5     &   0.00368          & 9157.16588$^f$    & 9157.166669      & $\pm0.00033 $        & I       &    0      &  0.00771      \\      
4405.36130$^a$    & 4405.362062    & $\pm0.00110 $    &   I     & -12946       &  -0.00076          & 9479.07808$^g$    & 9479.078860      & $\pm0.00030 $        & I       &  877      &  0.01807      \\      
5390.88365$^b$    & 5390.884412    & $\pm0.00219 $    &   I     & -10261       &  -0.00078          & 9479.26216$^g$    & 9479.262940      & $\pm0.00043 $        & II      &  877.5    &  0.01985          \\  
5391.06672$^b$    & 5391.067482    & $\pm0.00420 $    &   II    & -10260.5     &  -0.00124          & 9479.99632$^g$    & 9479.997100      & $\pm0.00034 $        & II      &  879.5    &  0.01991          \\  
6857.60585$^b$    & 6857.606632    & $\pm0.00197 $    &  I      & -6265        & -0.00013           & 9480.17764$^g$    & 9480.178420      & $\pm0.00029 $        & I       &  880      &  0.01771      \\      
6857.78939$^b$    & 6857.790172    & $\pm0.00136 $    &  II     & -6264.5      & -0.00011           & 9520.18901$^g$    & 9520.189789      & $\pm0.00037 $        & I       &  989      &  0.02091      \\      
7240.43348$^c$    & 7240.434280    & $\pm0.00087 $    &  I      & -5222        & -0.00296           & 9522.94074$^g$    & 9522.941519      & $\pm0.00046 $        & II      &  996.5    &  0.01978          \\  
7532.97111$^d$    & 7532.971914    & $\pm0.00212 $    &  I      & -4425        & -0.00213           & 9523.12551$^g$    & 9523.126289      & $\pm0.00034 $        & I       &  997      &  0.02103      \\      
7533.16062$^d$    & 7533.161424    & $\pm0.00126 $    &  II     & -4424.5      &  0.00385           & 9524.04374$^g$    & 9524.044519      & $\pm0.00044 $        & II      &  999.5    &  0.02164          \\  
8320.65776$^d$    & 8320.658570    & $\pm0.00080 $    &  I      & -2279      &  0.00072           &        &        &     &                                       &       &                 \\

\hline
\end{tabular}
\end{center}
\textbf
{\footnotesize Note }
(a) SuperWASP, (b) WISE, (c) CzeV, (d) ASAS-SN, (e) TESS, (f) NEXT, (g) WH60.
\end{table*}

\section{Photometric Solutions}
\begin{figure*}
\begin{center}
\begin{tabular}{c@{\hspace{0.3pc}}c}
\includegraphics[angle=0,scale=0.27]{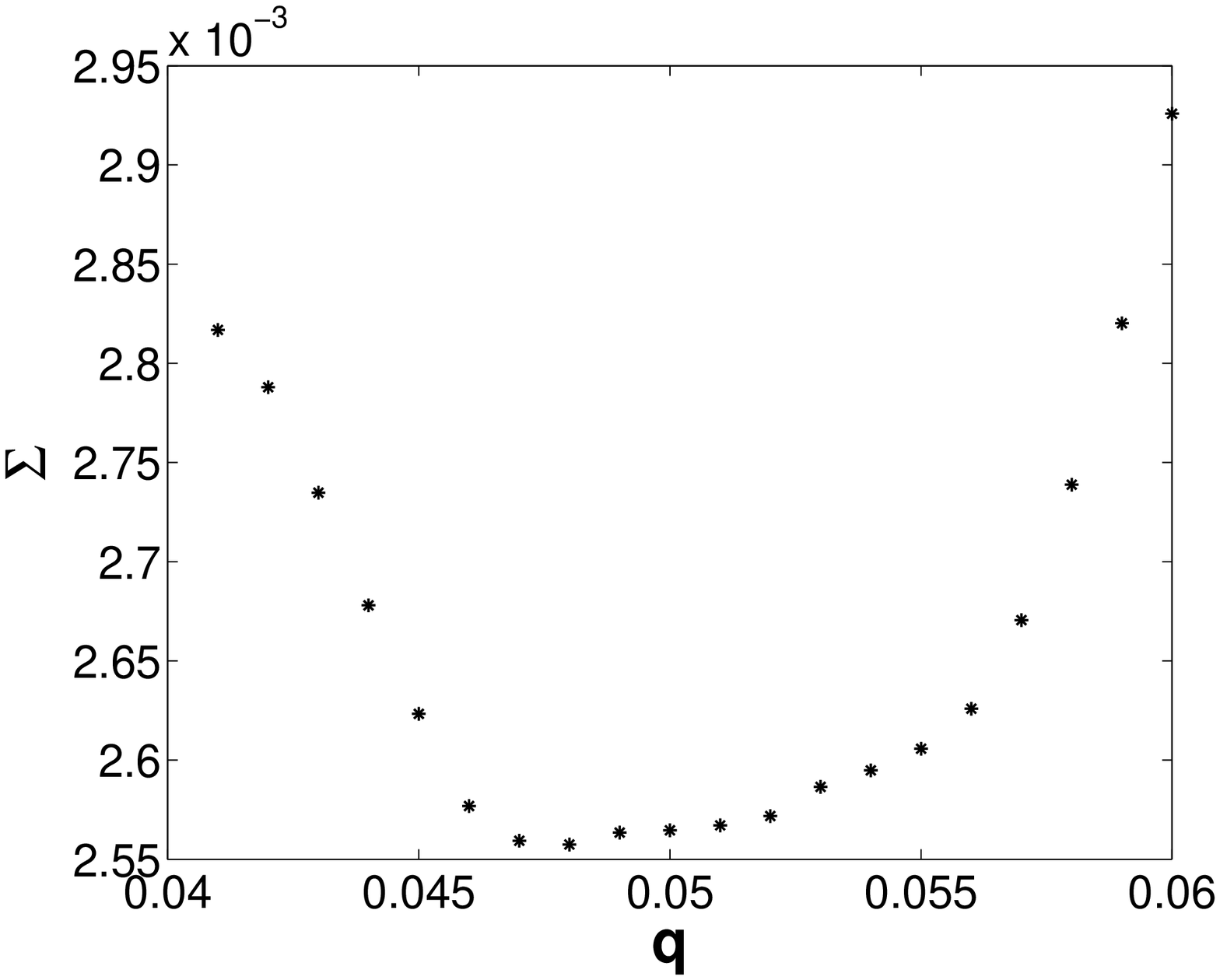}
\includegraphics[angle=0,scale=0.27]{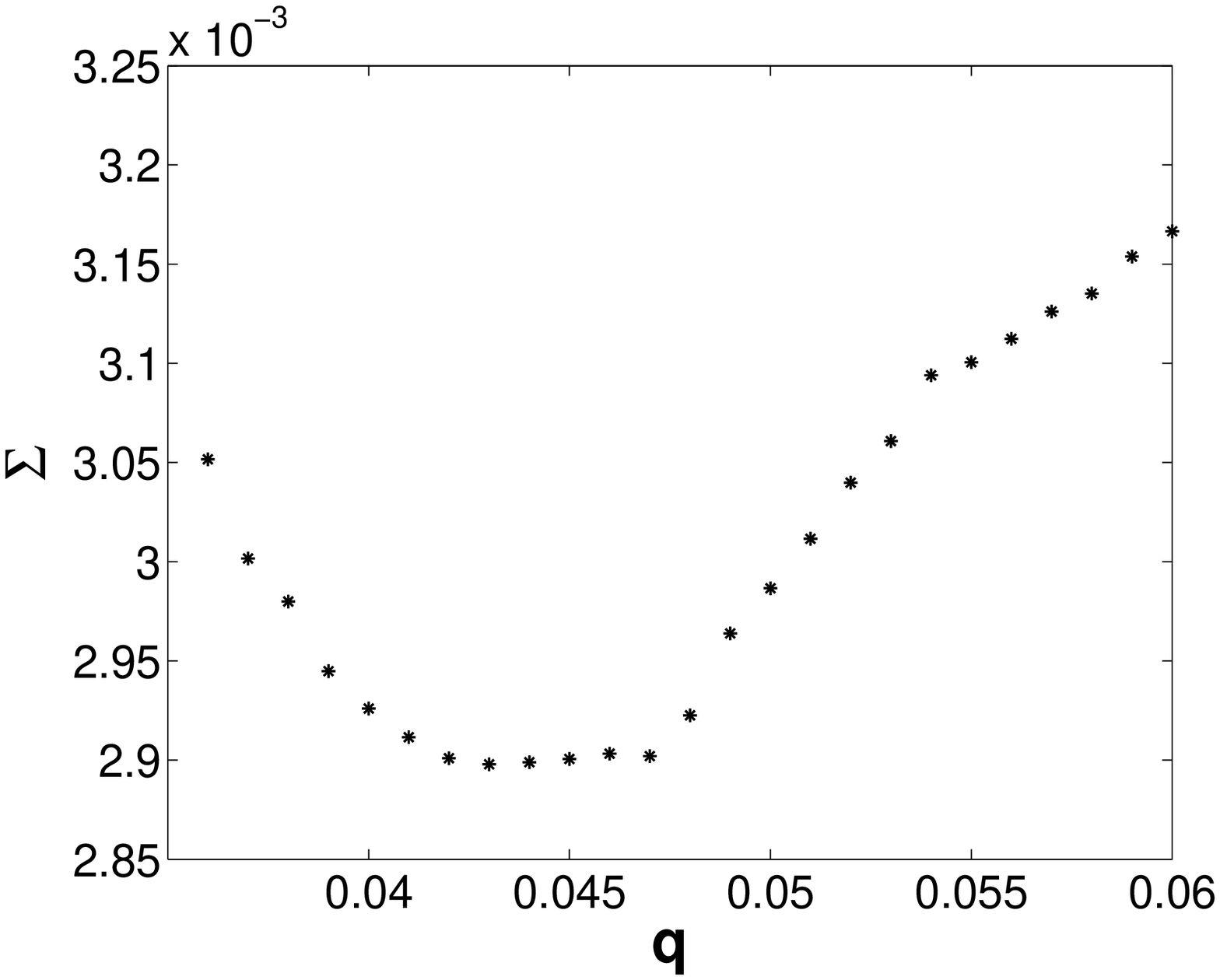}
\includegraphics[angle=0,scale=0.27]{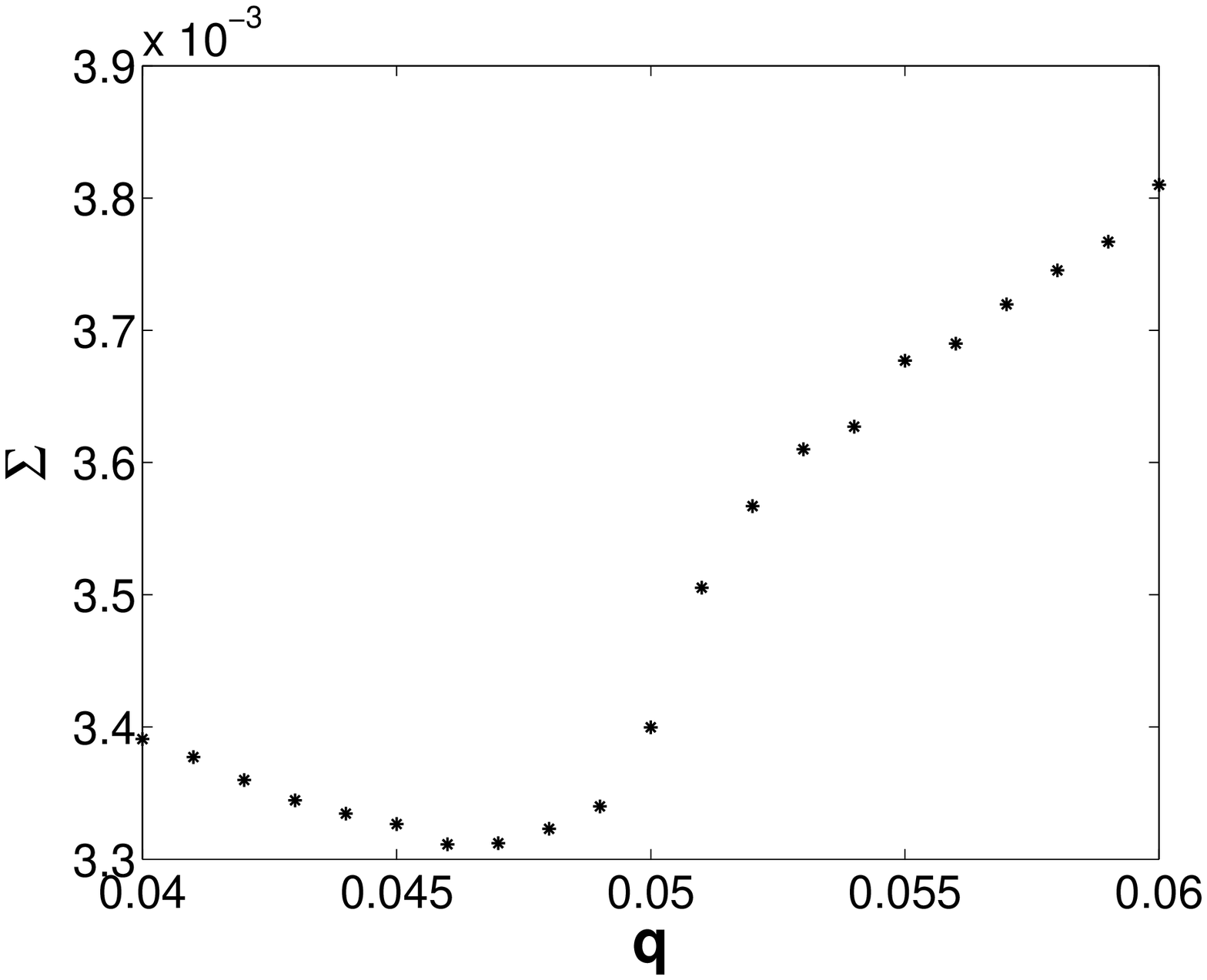}
\end{tabular}
\end{center}
\caption{Mass ratio q vs mean residuals $\sum$  for TYC 4002-2628-1 derived with the three light curves. The left panel, the middle panel and the right panel were derived from $LC_{1}$, $LC_{2}$ and $LC_{3}$, respectively. }
\label{fig-qsearch}
\end{figure*}

\begin{figure}
\begin{center}
\begin{tabular}{c@{\hspace{0.3pc}}c}
\includegraphics[width=8cm]{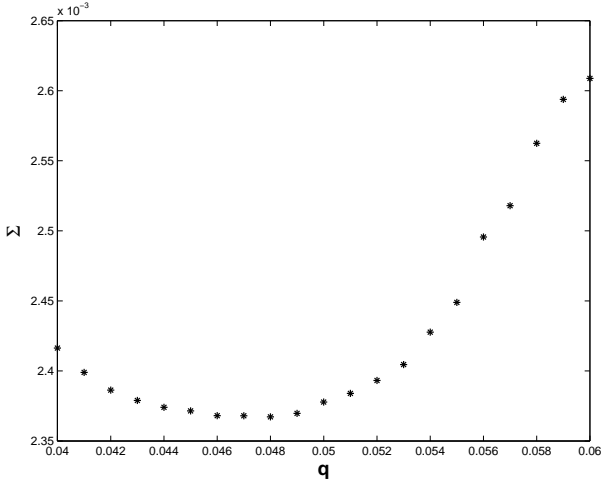}

\end{tabular}
\end{center}
\caption{Mass ratio q vs mean residuals $\sum$  for TYC 4002-2628-1 derived from the time series BVRI light curves. }
\label{fig_qsearch_all}
\end{figure}

\begin{table}
\begin{center}
\caption{Photometric solutions of TYC 4002-2628-111}
\label{solution_all}
\small
\begin{tabular}{ccc}
\hline\hline
Parameters                 &  Sol$_{BVRI}$     &  Sol$_{TESS}$ \\\hline
t$_{0}$                     &     9479.077975$^a$   &  1743.877075$^{b}$            \\
P$_{ZERO}$             &   0.3670584143          &  0.3670388842            \\
$T_{1}(K)$                    &      6032             &  6032            \\
$T_{2}(K)$                    &      6044(6)          &  6151(3)            \\
q($M_2/M_1$)                  &      0.0482(1)        &  0.0482(fixed)          \\
$i(^{\circ})$                 &      69.9(1)          &  69.7(1)             \\
$\Omega_{in}$                 &      1.782            &  1.782           \\
$\Omega_{out}$                &      1.750            &  1.750           \\
$\Omega_{1}=\Omega_{2}$       &      1.780(1)         &  1.771(1)         \\
$f$                           &      5(4)$\%$         &  35(2)$\%$         \\
$r_{1}$                       &      0.636(1)         &  0.641(1)          \\
$r_{2}$                       &      0.166(2)         &  0.173(1)          \\
$\Sigma{\omega(O-C)^2}$       &      0.002015         &  0.0000032          \\
\hline
\end{tabular}
\end{center}
\textbf
{\footnotesize Note }
(a) HJD-2450000, (b) BJD-2457000.
\end{table}

\begin{table}                                                                             
\begin{center}                                                                          
\caption{Luminosity ratio and spot parameters derived from the $BVRI$ light curves}\label{spot_solution}                                                                                          
\small                                                                                                                                                         
\begin{tabular}{lllllll}                                                                                                                                          
\hline\hline                                                                                                                                                   
Parameters                    &  Sol.1 ($LC_{1}$)                  &   Sol.2 ($LC_{2}$)             &   Sol.3 ($LC_{3}$)       \\\hline                       
$L_{1}/(L_{1}+L_{2}$)($B$)    &                                    &           0.9344(9)            &        0.9344(7)              \\                           
$L_{1}/(L_{1}+L_{2}$)($V$)    &      0.9348(9)                     &                                &                               \\                        
$L_{1}/(L_{1}+L_{2}$)($R$)    &      0.9349(6)                     &           0.9349(8)            &        0.9349(1)               \\                          
$L_{1}/(L_{1}+L_{2}$)($I$)    &      0.9350(5)                     &           0.9350(9)            &        0.9350(10)              \\                           
Colatitude(radian)            &      2.7449(401)                   &           1.8300(27)           &        0.4735(121)             \\                        
Longitude(radian)             &      0.4137(347)                   &           4.77498(317)          &        4.2472(223)            \\                        
Radius(radian)              &      0.7861(50)                    &           0.3011(101)          &        0.24632(121)            \\                      
$T_{s}/T$                     &      0.9269(72)                    &           0.9869(57)           &        0.92678(51)             \\                     
\hline                                                                                                                                                        
\end{tabular}                                                                                                                                                  
\end{center}                                                                                                                                               
\end{table}

The photometric solutions of TYC 4002-2628-1 were derived from the Wilson-Devinney (W-D) program of 2013 version \citep{Wilson1971,
Wilson1990,Van2007,Wilson2008,Wilson2010,Wilson2012}. The temperature $T_{eff}$ obtained by the spectral fitting is close to the temperature ($T_{B-V} = 5942 K$) derived from the $B - V$ color index after the correction of interstellar extinctions \citep{Green2018}.
TYC 4002-2628-1 is an extremely low mass ratio contact binary, the more massive star provides most of the luminosity to the whole system. Meanwhile, the target is a total eclipse at the secondary minimum, and the spectrum was observed at 0.55 phase. Therefore the temperature ($6032$ K) derived from spectral fitting was used as the temperature of the more massive star in the following analysis. The gravity darkening coefficients for the elements were fixed to be $g_1=g_2=0.32$ \citep{1967ZA.....65...89L} and the values of bolometric albedo coefficients $A_1=A_2=0.5$ \citep{1969AcA....19..245R}.  While the bandpass limb-darkening and the bolometric coefficients of the two stars were obtained from van Hamme's table \citep{1993AJ....106.2096V}.

Owing to lack of the radial velocity (RV) data, the $q$-search method was applied to get the initial mass ratio. The initial mass ratio can be obtained as follows: firstly, consecutive photometric solutions were calculated for a series of mass ratios, then the relationship between the tested mass ratios and the mean residuals were derived. For comparison, the results of the $q$-search method derived from the three set multi-color light curves were illustrated in Figure \ref{fig-qsearch}. As can be seen from Figure \ref{fig-qsearch}, the minimum values for $LC_{1}$, $LC_{2}$ and $LC_{3}$ are q=0.048, q=0.043 and q=0.047, respectively.  If the light curves of the binary obtained at different times remain unchanged, the results of $q$-search should be identical. Nevertheless, the light curves always varies with time, more or less, due to various reasons. Among them, the spot activities on the star surface is the most common reason.  So the results of the $q$-search method derived from the light curves are slightly different. At the same time, we also used the time series light curves to obtain the initial mass ratio. During the procession, the zero point of the orbital ephemeris ($t_{0}$), the orbital period ($P_{0}$), inclination (i), potential (Ω), temperature of the secondary ($T_{2}$) and luminosity of the primary
($L_{1}$) were the adjustable parameters. The results of the $q$-search method are displayed in Figure \ref{fig_qsearch_all}.  From this figure, we can see that the minimum mass ratio is the same as the one derived from $LC_{1}$, both of which are 0.048.  Since $LC_{1}$  observed by NEXT is more homogeneous than the other two ( $LC_{1}$ and  $LC_{2}$), the minimum value $q$=0.048 was set as the initial mass ratio and set as an adjustable parameter in the following photometric solutions.

\begin{figure*}
\begin{center}
\begin{tabular}{c@{\hspace{0.3pc}}c}
\includegraphics[angle=0,scale=0.27]{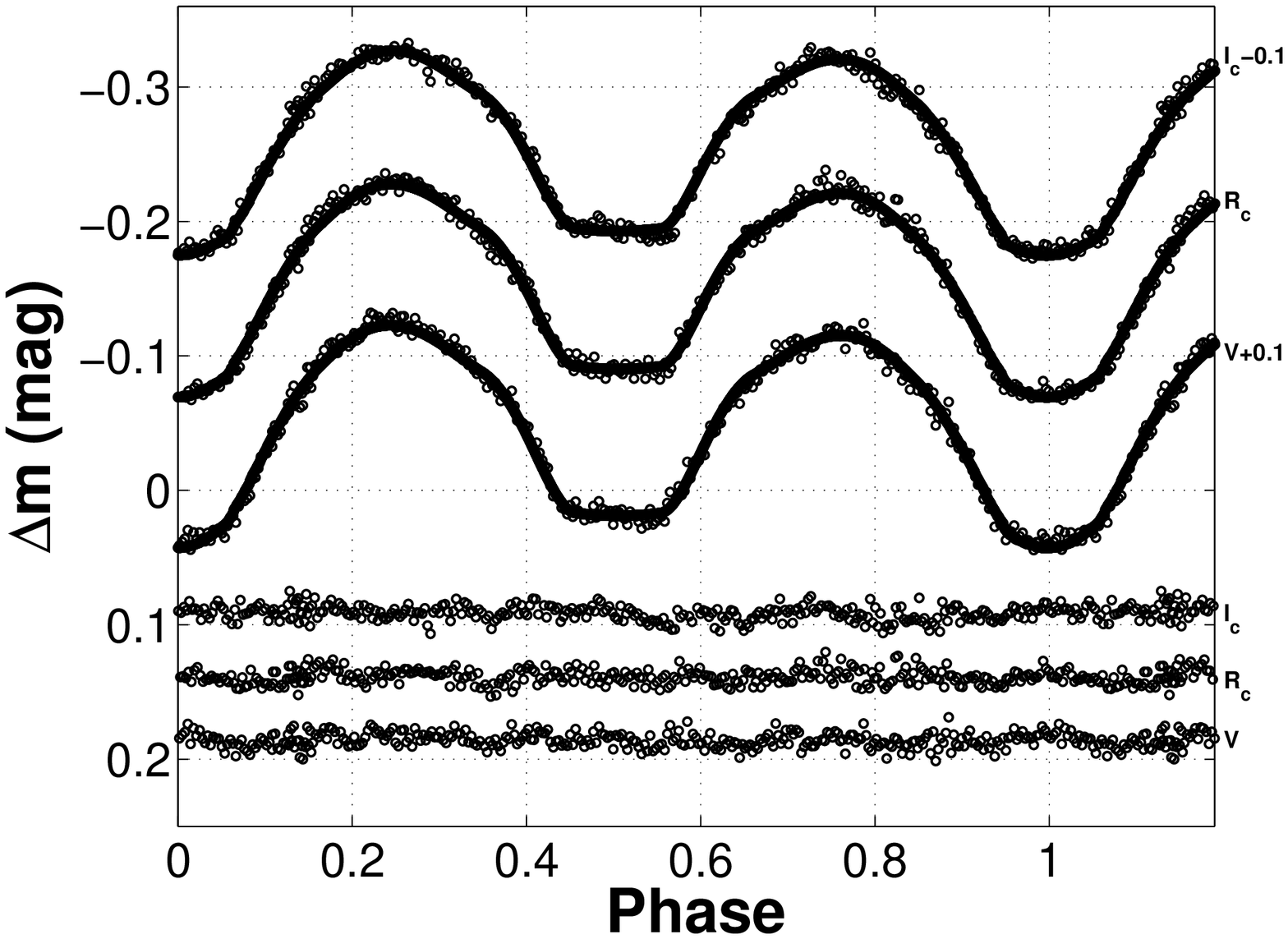}
\includegraphics[angle=0,scale=0.27]{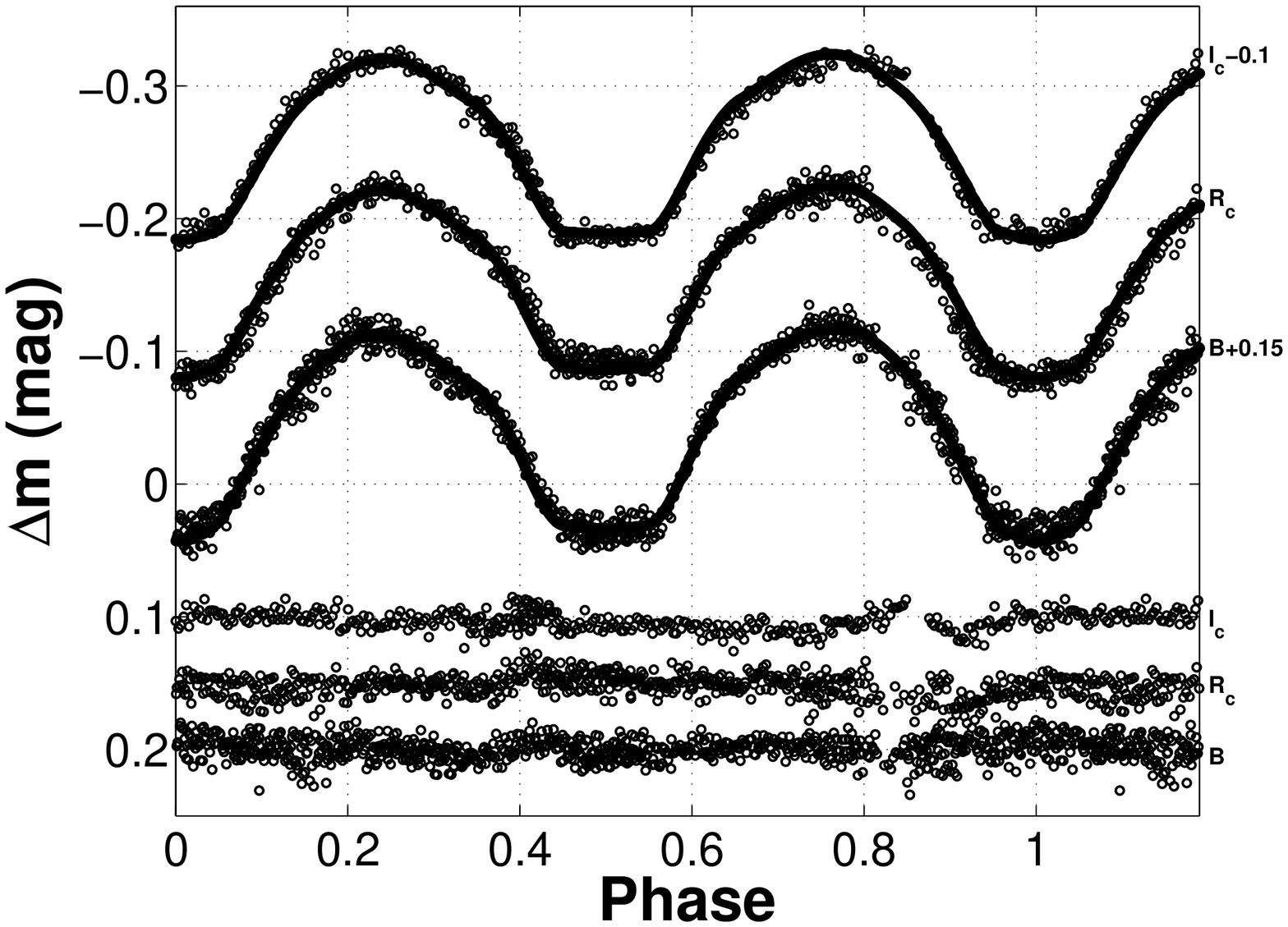}
\includegraphics[angle=0,scale=0.27]{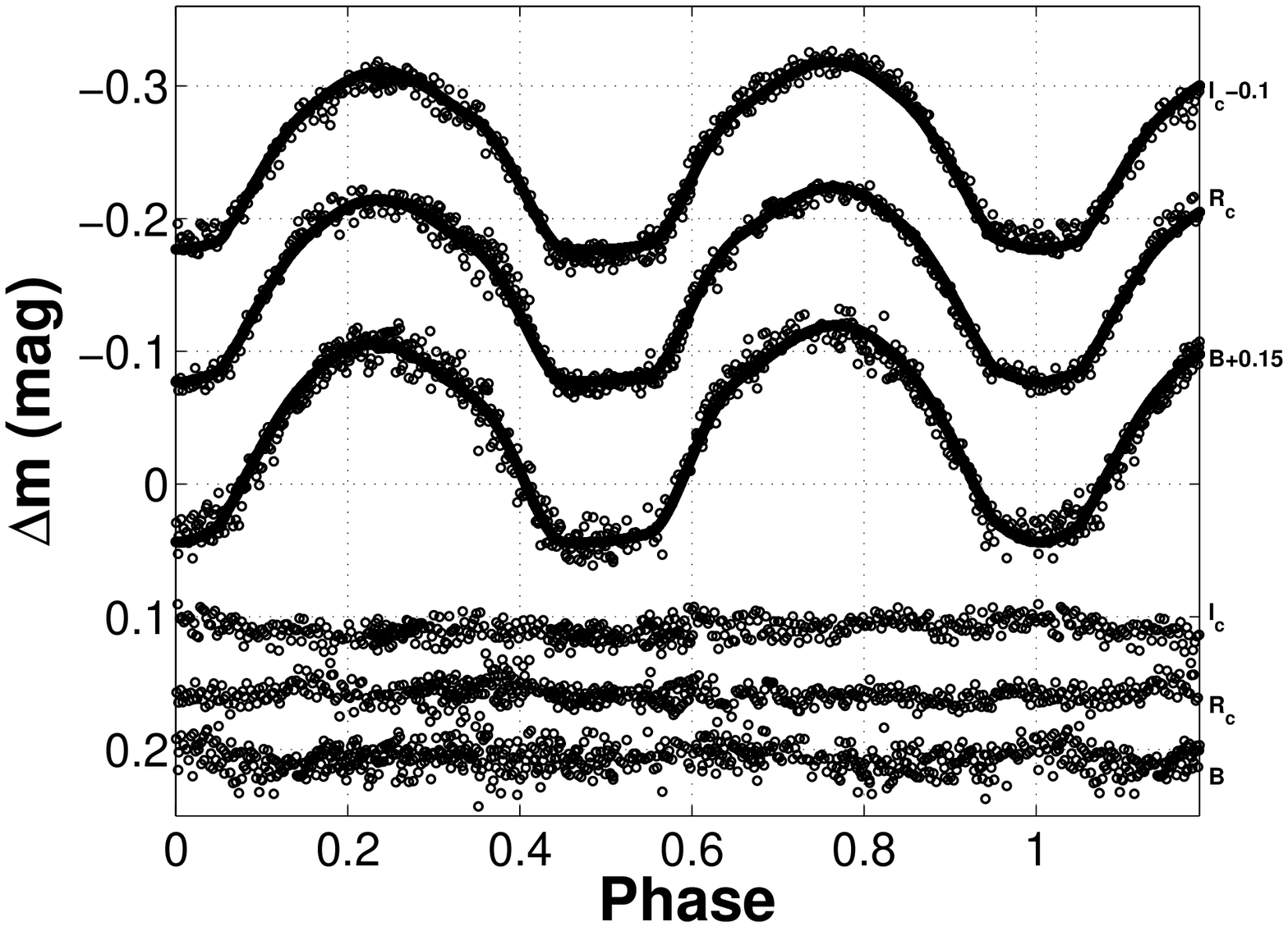}

\end{tabular}
\end{center}
\caption{The theoretical light curves of TYC 4002-2628-1. The residuals are plotted at bottom of the corresponding panels. The left panel observed in 2020 November ({$LC_{1}$}), the middle panel observed in 2020 September and October ({$LC_{2}$}), while the right panel obtained in 2021 November ({$LC_{3}$}).}
\label{fig_all_fit}
\end{figure*}

To generate a unique set of orbits and binary parameters consistent with each curve, we solved the multi-band time series light curves simultaneously using the 2013 version of the Wilson-Devinney program. As displayed in Figures \ref{fig-all} and \ref{fig-compare}, the light curves are asymmetric, which may be due to magnetic activities, so a star spot is added in each corresponding interval of time. The time spans corresponding to the three sets of light curves ($LC_{1}$, $LC_{2}$ and $LC_{3}$ ) are used to define the active time of each spot in the subsequent photometric solution. Each spot is set to be active only for the corresponding time span, and spot parameters are adjusted to alternate between the three spots. After some interaction, the final solutions for the light curves were derived and listed in Table \ref{solution_all} as Sol$_{BVRI}$, and the spot parameters are shown in  Table \ref{spot_solution}. 
The fitted light curves derived from the photometric solutions are displayed as solid lines in Figure \ref{fig_all_fit}, in which residuals (observed minus calculated light curves) from
the solutions are shown in black dots. 
Since the third light will affect the mass ratio of photometric solution, We try to check the existence of the third
body via adding a third light during the photometric analysis. However,
the value of third light is always negative.  We also checked the Gaia website and found that the closest star to the target is about 6 arcsecond away, which is larger than the photometric aperture. 
At the same time, the magnitude of nearest star is 20.6 magnitude, much dimmer than our target (11.3, g mag). Therefore, the contribution of the third light can be ignored.

\begin{figure*}
\begin{center}
\begin{tabular}{c@{\hspace{0.3pc}}c}

\includegraphics[angle=0,scale=0.27]{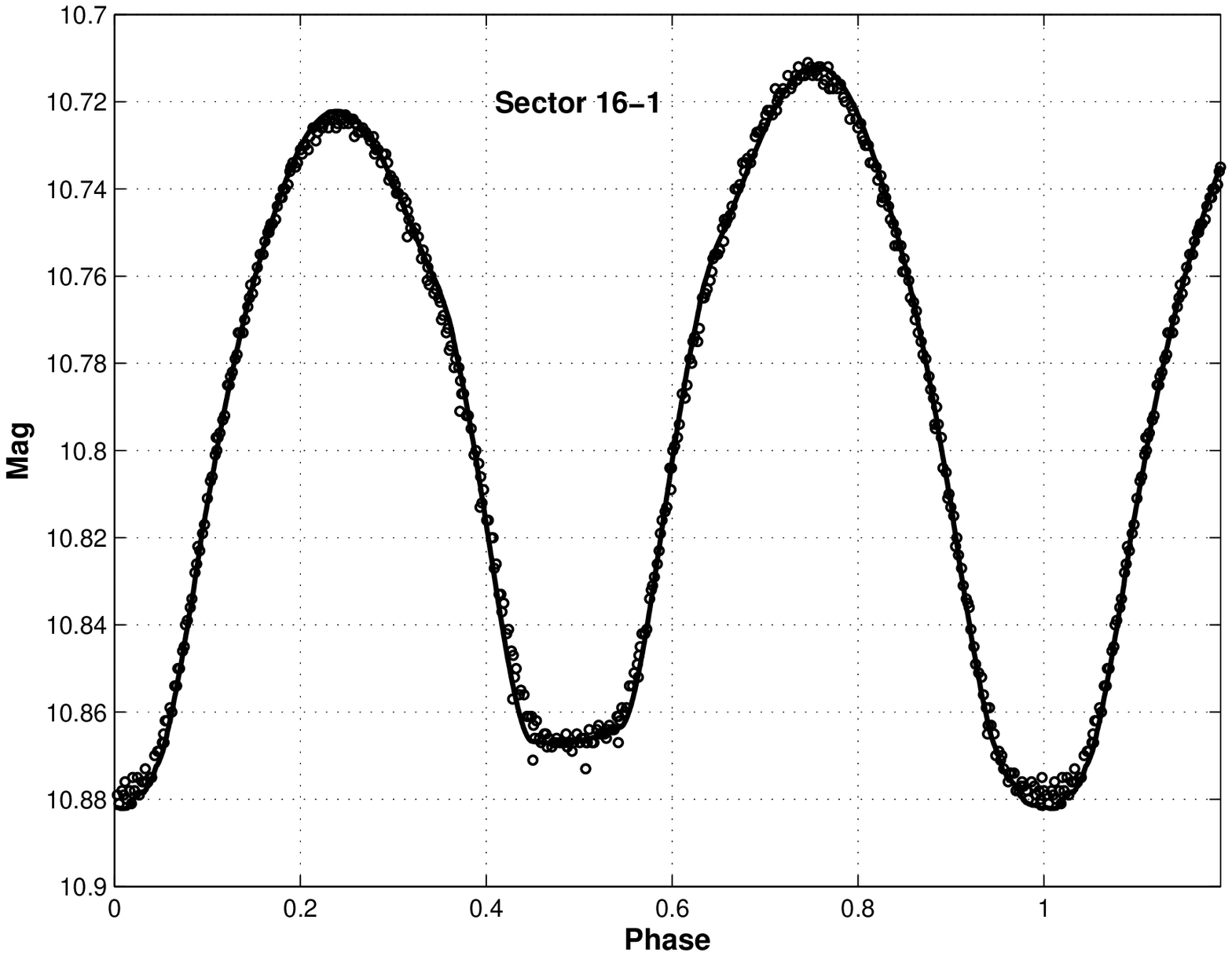}
\includegraphics[angle=0,scale=0.27]{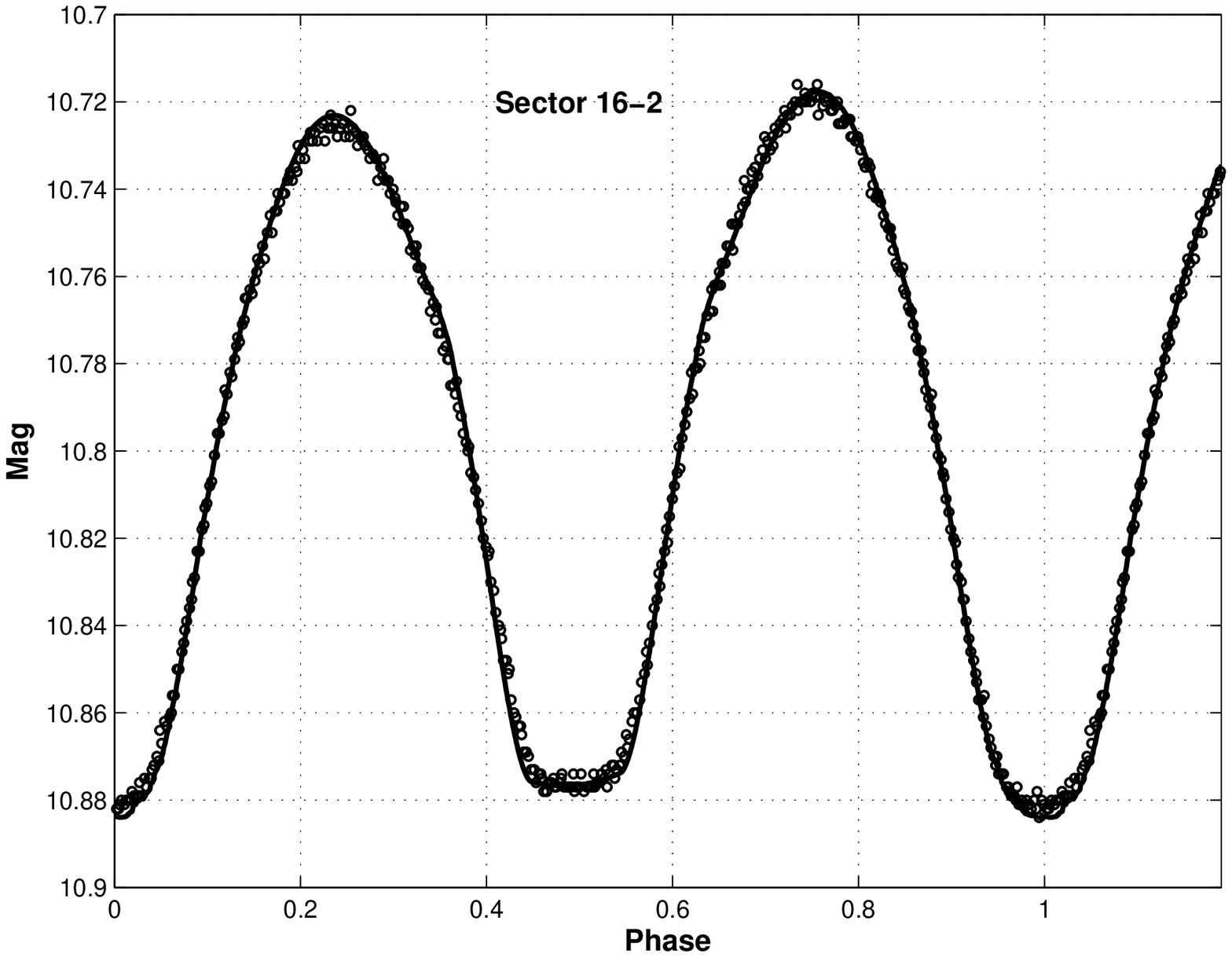}
\includegraphics[angle=0,scale=0.27]{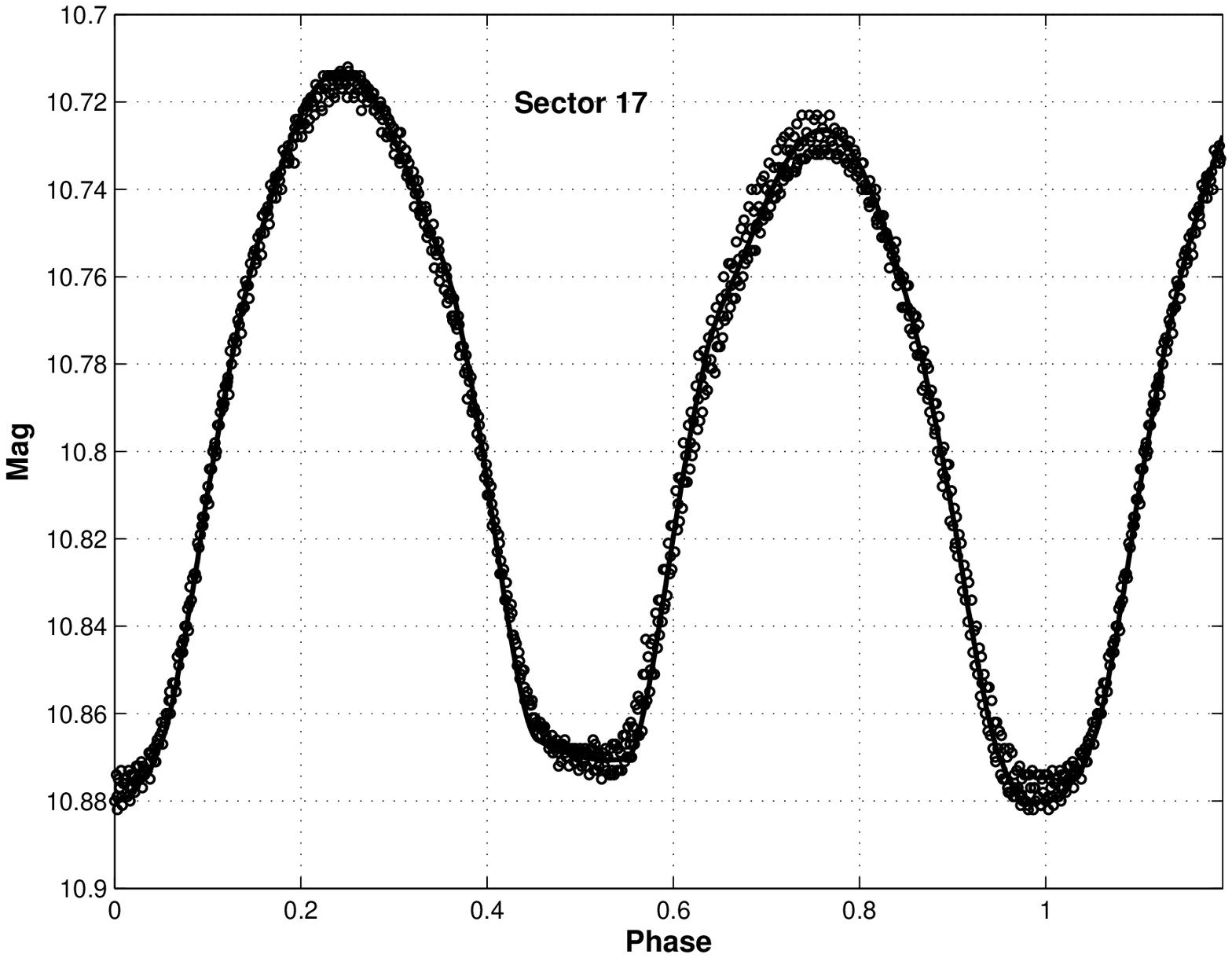}\\
\includegraphics[angle=0,scale=0.27]{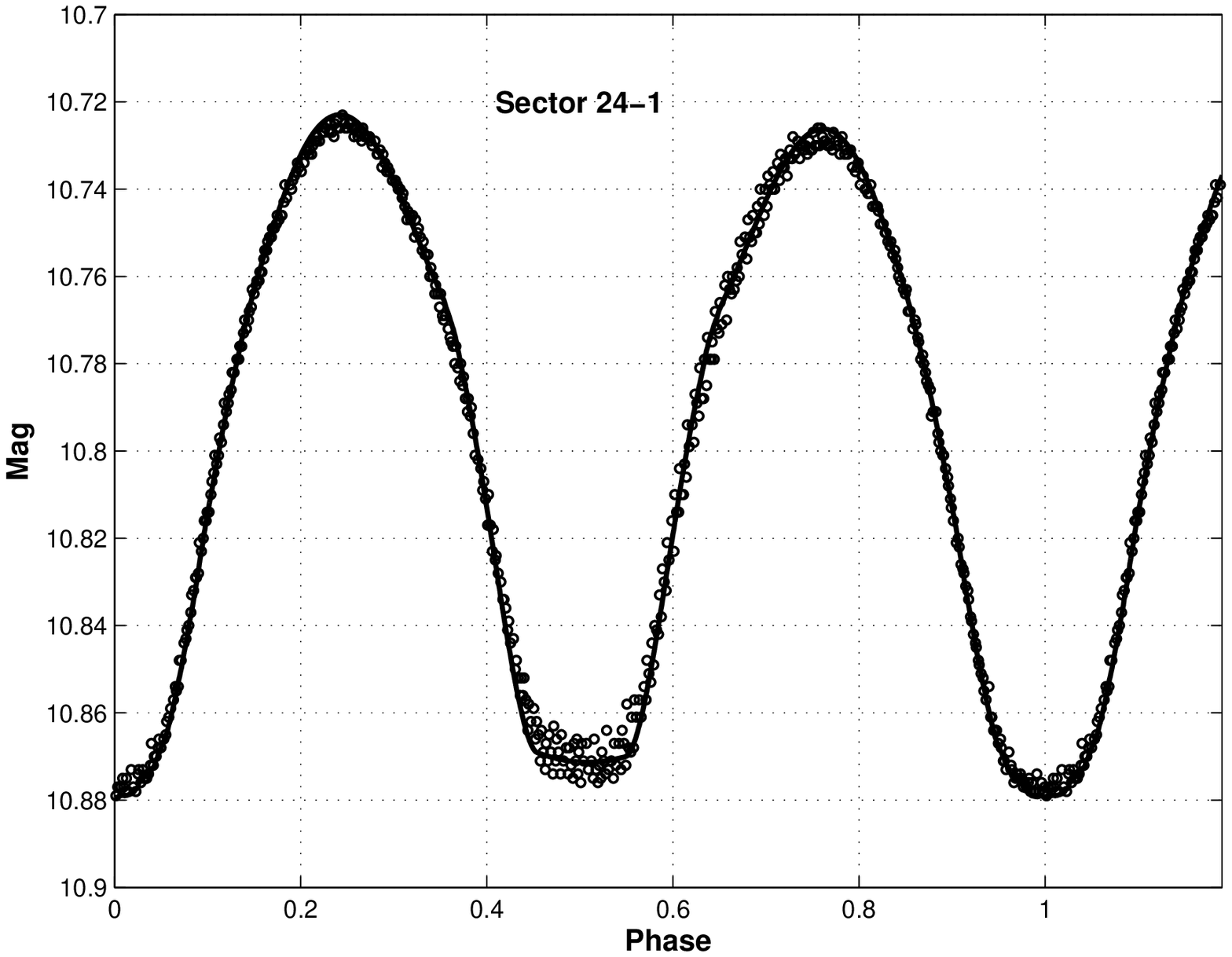}
\includegraphics[angle=0,scale=0.27]{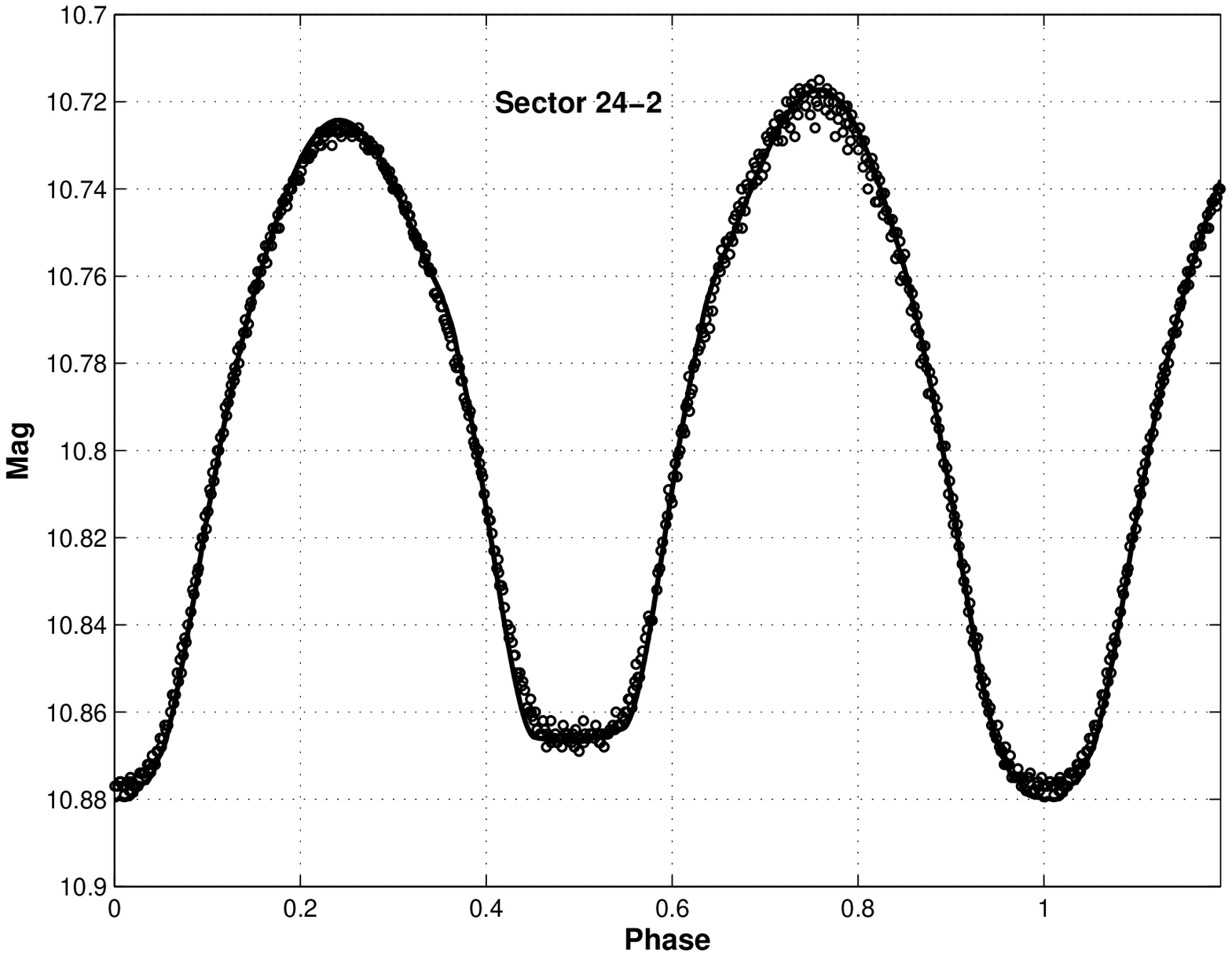}\\

\end{tabular}
\end{center}
\caption{ Light curves of TYC 4002-2628-1 obtained by TESS. The solid lines represent the theoretical light curves.}
\label{fig-tess}
\end{figure*}

Based on the shapes of the light curves, five sets of light curves were obtained from the TESS data (Sectors 16, 17, and 24), which are shown in Figure \ref{fig-tess}. The time spans corresponding to these five sets of light curves are used to define the active time of each spot in the subsequent photometric solution. Considering the TESS data for TYC 4002-2628-1 are in a 30-minute cadence and the short period of the target, the phase smearing effect inevitably affects the photometric solutions, such as the contact degree, orbital inclination and other parameters \citep{zola2017}. In order to reduce the influence of the phase smear effect, the control parameter NGA$=$3 (Number of Gaussian quadrature abscissa in phase or time smearing simulation) was used \citep{zola2017}.
Using the same method as above, the final photometric solutions for the time series light curves were obtained and exhibited in Table \ref{solution_all} as Sol$_{TESS}$ and Table \ref{phsolutions_tess}.

\begin{table*}
\begin{center}
\caption{\textbf{Luminosity ratio and spot parameters derived from the data of TESS}}
\label{phsolutions_tess}
\small
\begin{tabular}{lllllllll}
\hline\hline
Parameters                   & Sector16-1       & Sector16-2             & Sector17                  &  Sector24-1              & Sector24-2   \\\hline
$L_{1}/(L_{1}+L_{2}$)        & 0.9284(1)         & 0.9284(1)             & 0.9284(1)      & 0.9284(1)               & 0.9284(1)      \\
Colatitude(radian)           & 0.0646(102)       & 0.0250(11)            & 0.0718(132)    &  0.1511(102)            &   0.1212(53)    \\       
Longitude(radian)            & 5.5173(133)       & 5.2820(452)           & 1.1281(136)    &  0.5388(136)            &   5.7113(195)    \\      
Radius(radian)             & 0.7266(19)        & 0.8122(fixed)         & 0.5038(33)     &  0.5725(9)              &   0.5315(6)      \\  
$T_{s}/T$                    & 0.8806(8)         & 0.8773(fixed)         & 0.8114(7)      &  0.9691(9)              &   0.9150(6)      \\  

\hline
\end{tabular}
\end{center}
\end{table*}

\section{Discussions and Conclusions}

Based on our observations and TESS data, eight complete light curves of TYC 4002-2628-1 were obtained. The light curves show day to day, month to month and year to year variations, indicating strong spot activities. As showed by the left panel of Figure \ref{fig-compare}, the secondary flat minimum (0.5 phase) is deeper than the primary minimum in $LC_{3}$. This phenomenon implies that the binary system has changed its type (between W-type and A-type) in no more than one month. 
The light curves also show the reversal of O'Connell effect: Obvious positive O'Connell effects are found in $LC_{1}$, $LC_{2}$, Sector 17, and Sector 24-1, while other light curves show obvious negative O'Connell effect. Our photometric solutions indicate that the quick reversal of O'Connell effect can be explained by spot variation. All these phenomena indicate the magnetic activity is very active, which may be due to the deep convective envelope along with rapid rotation.

The photometric solutions derived from the multi-band time series light curves indicate that TYC 4002-2628-1 is   an extremely mass ratio contact binary, with a shallow contact degree of $f=(\Omega_{in}-\Omega)/(\Omega_{in}-\Omega_{out})=5\%$, where $\Omega_{in}$ is the potential when the Roche lobe arriving at Lagrange $L_1$ point of the system and $\Omega_{out}$ is the maximum potential of the Roche lobe (arriving at $L_2$) and $\Omega$ is the stars's actual potential. The photometric solutions derived from different light curves are slightly different, which may be caused by the spot activity and the phase smearing effect.
Supposing that the more massive component of the
binary is a main-sequence star, we can estimate the mass of M$_1$ as
 1.11 M$_\odot$,  according to the online table\footnote{http://www.pas.rochester.edu/$\sim$emamajek/EEM$\_$dwarf$\_$UBVIJHK$\_$
colors$\_$Teff.txt} \citep{Pecaut2013}. The less massive component of the target can be calculated as M$_2$ = $q \times$ M$_1$ = 0.0535 M$_\odot$.  According to
Kepler's third law ($M=0.0134a^3/P^2$), the
distances between the two stars can be estimated as a = 2.27R$_\odot$. Based on the photometric solutions of $LC_1$, the radii of the primary and secondary components can be calculated as  R$_1$ =$r_1 \cdot a$ = 1.45R$_\odot$, R$_2$ = $r_2 \cdot a$ = 0.39R$_\odot$. The luminosity of L$_1$ = 2.49L$_\odot$ and L$_2$ =0.16L$_\odot$ were derived from the Stefan-Boltmann's law (L=$4\pi\sigma$$T^4R^2$).

\subsection{The period changes}

 By analyzing all of the calculated eclipsing minima, the orbital period variation of TYC 4002-2628-1 can be divided into two parts (see Figure \ref{oc}). The first part of the $O-C$ diagram reveals that the period remains unchanged, which indicate there was no mass exchange between the two components. The second part of the $O-C$ diagram shows a parabolic variation, which implies its period has changed suddenly. The discontinuity of the period may be caused by a sudden mass transfer between the two components.
 The quadratic term in Equation \ref{oc3} implies that the period is long-term increasing at a rate of $dP/dt=1.62\times{10^{-5}}day\cdot yr^{-1}$ (1.40 s yr$^{-1}$ ).  If
this is the case, TYC 4002-2628-1 would be the contact binary with the highest orbital period increasing rate so far. The long-term period increase can be explained by the mass transfer from less massive component to the more massive one. The rate of mass transfer can be derived from the formula,
\begin{eqnarray}
{\dot{P}\over P}=-3\dot{M_1}({1\over M_1}-{1\over M_2}),
\end{eqnarray}
 by supposing the mass transfer is conservative.
The decreasing rate for the less massive star is computed to be $dM_2/dt=8.25\times10^{-7}\,M_\odot$ yr$^{-1}$.
Considering the observation span for the parabolic variation of $O-C$ diagram is less than three years, the continuous period increase maybe just a part of the periodic variation. More eclipsing minima are required to test this possibility.

\subsection{Potential of merger}

According to the theory of Darwins instability \citep{Hut1980}, if $J_{spin} \over J_{orb}$ $>$ 1/3, the contact binaries will move towards merger, where $J_{spin}$ and $J_{orb}$ represent the spin angular momentum and the orbital angular momentum, respectively.  The value of $J_{spin} \over J_{orb}$ can be derived from the following equation \citep{Yang2015}:
\begin{equation}
{J_{spin} \over  J_{orb}} = {1+q \over q }[(k_{1}r_{1})^2 + (k_{2}r_{2})^2q],
\label{jspin}
\end{equation}
where $k_{1}$ and $k_{2}$ are the dimensionless gyration radii of the corresponding components, $r_{1}$ and $r_{2}$ are the relative radii of the components.
The value of  $k_{1}$ was calculated using the equation: $k_{1} = -0.250M + 0.539$ \citep{Landin2009,Christopoulou2022}. Due to the very low mass of the less massive component, the value of $k_{2}^2$ was fixed as 0.205, which is consistent with a fully convective star \citep{Arbutina2007}. By substituting the above values into Equation \ref{jspin}, we can get  $J_{spin}\over J_{orb}$ $=$ 0.6210, which is more than 1/3.
For comparison, $k_{1}^2=k_{2}^2=0.06$ \citep{Rasio1995a,Li2006} was adopted to recalculate $J_{spin}\over J_{orb}$. Then the value of $J_{spin}\over J_{orb}$ can be determined to be 0.5380, which is still more than 1/3.

For further investigation of the stability of TYC 4002-2628-1,
 we have used the methods introduced by
\cite{Wadhwa2021} to estimate the instability parameters. According to the mass of primary component ($M_{1}$) and the fill-out factor ($f$), the instability mass ratio $q_{inst} $ was calculated as 0.086, which is larger than the mass ratio ($q\sim0.0482$) derived from WD code.  In order to estimate the instability separation ($A_{inst}$), $k_{1}$ was calculated using the equation: $k_{1} = -0.250M + 0.539$ \citep{Landin2009,Christopoulou2022} and the value of $k_{2}^2$ was fixed to be 0.205 as mentioned above. Then $A_{inst}=3.034 R_\odot$ was determined. Both of the instability parameters ($q_{inst}$, $A_{inst}$) are larger than the actual parameters ($q=0.0482$, $A = 2.27R$$_\odot$), indicating that the target is a potential merger candidate.

 analyzing the light minimum times, a highest rate of orbital period increase ($dP/dt=1.62\times{10^{-5}}day\cdot yr^{-1}$ = 1.40 $s \cdot yr^{-1}$ ) was obtained. However, due to the short time span of observation,
more eclipsing minima are needed to verify its authenticity. By studying the ratio of spin angular momentum to orbital angular momentum ($J_{spin} \over J_{orb} $$> 1/3$), the instability mass ratio ($q_{inst}> q$)

In summary, the first multi-band photometric and spectral observations of the total eclipse binary TYC 4002-2628-1 were presented. The photometric solutions indicate the system is an extremely low mass ratio ($q\sim0.0482$) with a shallow contact degree. So far, there is only one similar system with a mass ratio less than 0.05: V1187 Herculis. Searching for more such kind binaries will help us to test and constrain the merging theory.
The $O-C$ diagram shows that the period of TYC 4002-2628-1 has changed suddenly  $BJD  2458743.$ By  and the instability separation ($A_{inst}> A$), we found that the system, like ZZ PsA, SX CrV, ASAS J165139+2255.7 and V1187 Her, is a candidate of merging binary.

\section*{Acknowledgements}
\addcontentsline{toc}{section}{Acknowledgements}

This work is supported by the National Natural Science Foundation of China (NSFC) (No.12063005), and the Joint Research Fund in Astronomy (No. U1931103) under cooperative agreement between NSFC and Chinese Academy of Sciences (CAS), and by the Qilu Young Researcher Project of Shandong University, and by the Chinese Academy of Sciences Interdisciplinary Innovation Team, and by the Cultivation Project for LAMOST Scientific Payoff and Research Achievement of CAMS-CAS, and
the Program for Innovative Research Team (in Science and Technology) in the University of Yunnan Province (IRTSTYN), and Yunnan Local Colleges Applied Basic Research Projects (2019FH001-12).  The calculations in this work were carried out at Supercomputing Center of Shandong University, Weihai. We acknowledge the support of the staff of the Xinglong 2.16m telescope. This work was partially supported by the Open Project Program of the CAS Key Laboratory of Optical Astronomy, National Astronomical Observatories, ChineseAcademy of Sciences. Many thanks to anonymous reviewer for their very helpful comments and suggestions, which greatly improved our manuscript.

This work makes use of  data collected by the TESS mission which are funded by NASA Science Mission directorate. We acknowledge the TESS team for its support of this work. This paper also used the data from ASAS-SN, which is funded in part by the Alfred P. Sloan Foundation under grant G202114192. This paper makes use of data from the DR1 of the WASP data (Butters et al. 2010) as provided by the WASP consortium,
and the computing and storage facilities at the CERIT Scientific Cloud, reg. no. CZ.1.05/3.2.00/08.0144
which is operated by Masaryk University, Czech Republic.


\section*{Data Availability}

The photometric data for the NEXT and WH60 used in this article are available in online
supplementary material.
The TESS data are publicly available at http://archive.stsci.edu/te
ss/bulk downloads.html, SuperWASP data are publicly available
at https://wasp.cerit-sc.cz/search and ASAS-SN data are publicly
available at https://asas-sn.osu.edu/variables/lookup. The WISE
data used in this paper are publicly available at http://variables.cn:88/wise/. The CzeV data used in this paper are derived from http://var2.astro.cz/czev.php?id=710.







\bsp	
\label{lastpage}
\end{document}